\documentclass[pdflatex,sn-mathphys-ay]{sn-jnl}

\usepackage{graphicx}%
\usepackage{multirow}%
\usepackage{amsmath,amssymb,amsfonts}%
\usepackage{amsthm}%
\usepackage{mathrsfs}%
\usepackage[title]{appendix}%
\usepackage{xcolor}%
\usepackage{textcomp}%
\usepackage{manyfoot}%
\usepackage{booktabs}%
\usepackage{algorithm}%
\usepackage{algorithmicx}%
\usepackage{algpseudocode}%
\usepackage{listings}%

\theoremstyle{thmstyleone}%
\theoremstyle{thmstyletwo}%

\theoremstyle{thmstylethree}%

\begin{document}

\title[Article Title]{Introduction to Strong Alfv\'enic MHD Turbulence}

\author{\fnm{Jungyeon} \sur{Cho}}\email{jcho@cnu.ac.kr}
\affil{\orgdiv{Department of Astronomy and Space Science}, \orgname{Chungnam National University}, \orgaddress{\street{99 Daehak-ro}, \city{Daejeon}, \postcode{34134}, \state{}\country{Korea (ROK)}}}







\abstract{ 
 Many astrophysical fluids are magnetized and turbulent. 
  Such fluids can be often described by magnetohydrodynamics (MHD).
In this review, we mainly consider MHD turbulence with a strong mean magnetic field
 whose energy density is greater than or equal to the local
 kinetic energy density.
In these fluids, the MHD waves, especially Alfv\'en waves,
 play a dominant dynamical role.
Alfv\'en waves travel along magnetic field lines and collisions of opposite-traveling Alfv\'en wave packets are essential for turbulence cascade.
We focus on strong turbulence regime, where nonlinear interaction during the collision is sufficiently strong and thus one collision is enough to complete turbulence cascade.
We will cover the following types of turbulence.
First, we review strong Alfv\'enic MHD turbulence.
If the mean magnetic field is very strong, wave packets move very fast and duration of collision is too short to complete turbulence cascade. 
Even in this case we will show that strong turbulence regime emerges on a small scale.
 Second, we will consider small-scale MHD turbulence, where interaction of small-scale variants of Alfv\'en waves (i.e., whistler waves) is
 important. 
Third, we review scaling relations in strong relativistic force-free MHD turbulence, where interaction of relativistic Alfv\'en waves is important.
Finally, we briefly discuss scaling relations in compressible MHD turbulence, where interaction of Alfv\'en waves is still important.
 }




\maketitle


\section{Introduction}\label{sec1}
 Since astrophysical fluids exhibit turbulent motions and magnetic fields are 
undoubtedly found in such fluids, magnetohydrodynamic (MHD) turbulence is 
an important field of study in astronomy. In most astrophysical plasmas, the 
magnetic Reynolds number easily exceeds $10^{10}$. In such a system, the usual 
expectation is that magnetic field is frozen into the fluid. The velocity field 
advects and stretches magnetic field lines and magnetic field exerts pressure 
and tension forces on velocity field. Therefore, MHD turbulence is 
different from pure hydrodynamic turbulence.

Consider incompressible MHD turbulence in a system threaded by 
a uniform external (or `mean') magnetic field. 
Then we may consider two extreme cases: strong mean field cases and weak/zero mean field cases. 
Numerical studies show that zero mean field cases are virtually identical to weak mean field cases 
after turbulence reaches a statistically stationary state
(see, for example,  \citealt{Cho14}). 
Therefore we may put weak and zero mean field cases into the same category.
When the mean field is weak/zero, turbulence motions can efficiently amplify magnetic field \citep[see, for example, ][]{Bat50,Kaz68, Meneg81, Kida91,Cho00dynamo,Schek04, ChoV09,Brand12}.
As a result of this turbulence dynamo process, magnetic energy density initially grows and, as magnetic field gets stronger,
magnetic back reaction becomes important.
When magnetic field becomes sufficiently strong,
turbulence dynamo process saturates due to magnetic back reaction 
and growth of magnetic energy density stops.
Numerical simulations \citep{Cho00dynamo,ChoV09} show that,
after turbulence reaches the saturation stage, coherence length of the magnetic field is comparable to the driving scale and that,
below a scale slightly smaller than the energy injection scale,
magnetic and kinetic energy densities are roughly in equipartition.
This result implies that, even if
the mean field is weak/zero, 
 turbulence on small scales can be regarded as turbulence threaded by a strong mean field.
 In this paper, we focus on MHD turbulence threaded by a strong mean field.

When external mean magnetic field is strong, MHD waves play important roles.
In incompressible fluids, 
there are two types of MHD waves: shear Alfv\'en and pseudo-Alfv\'en waves.
They have the same dispersion relation and propagate along the magnetic field
 lines at the Alfv\'en speed, which is proportional to the strength of the magnetic field: 
\begin{equation}
   V_A = B_0/\sqrt{4 \pi \rho},
\end{equation}
 where $B_0$ is the strength of the mean field and $\rho$ is density\footnote{
 In numerical simulations, it is customary to use a system of units in which $4 \pi $ does not appear (i.e., $4 \pi =1$). It is also customary to set $\rho=1$ in many (incompressible) simulations.
Therefore, $4 \pi \rho=1$ in many simulations.
If this is the case, we can use $V_A$ and $B_0$ interchangeably for the Alfv\'en speed. 
Note that, in the units, kinetic energy density ($\rho v^2/2$) 
and magnetic energy density ($B^2/8 \pi$)
become $v^2/2$ and $B^2/2$, respectively.
Hereinafter, we will call the system of units in which $4 \pi \rho=1$ `numerical' units.
 }.
 Detailed studies show that pseudo-Alfv\'en waves are only passively evolved by
 Alfv\'en waves \citep{GS95,GS97}.
In this sense, Alfv\'{e}n waves play fundamental roles in magnetized plasmas: they determine dynamics of 
incompressible magnetized plasmas with a strong background field.
Therefore, in this paper, we will focus on Alfv\'en waves.
It is important to note that Alfv\'en waves are non-dispersive and thus Alfv\'en waves moving in one direction do not interact. Only counter-traveling Alfv\'en waves interact.

When two opposite-traveling Alfv\'en wave packets of similar sizes collide, nonlinear interactions will distort themselves and give rise to energy cascade to smaller scales.
Then the question is the strength of the nonlinear interaction.
If nonlinear interaction is sufficiently strong, then one collision can be enough to complete energy cascade.
This regime of turbulence is called `strong' MHD turbulence.
On the other hand, if the strength of mean magnetic field is very strong, wave packets move very fast and thus duration of
interaction will be very short.
In this case, there will not be enough time to complete cascade.
This type of turbulence is called `weak' MHD turbulence.
In this paper, we focus on strong MHD turbulence.

Energy spectrum and anisotropy are of great importance in MHD turbulence.
Here, anisotropy means elongation of turbulence eddies (or wave packets).
In incompressible hydrodynamic turbulence, the famous Kolmogorov theory \citep{Kol41,Kol62} assumes isotropy of eddy shapes and predicts 
\begin{equation}
   E_v(k) \propto k^{-5/3},
\end{equation}
where
$E_v(k)$ is the kinetic energy spectrum.
In incompressible MHD turbulence, 
 \cite{Irosh64} and \cite{Kraich65} assumed isotropy and derived $k^{-3/2}$ spectra for both velocity and magnetic field.
 However, the assumption of isotropy has received a lot of criticism 
 \citep{Montg81,Sheba83, SG94, Montg95,Matt98prl}.
 In fact, it is natural for a dynamically important (i.e., strong) mean magnetic field to break a symmetry and provide a direction for anisotropy.
 Finally, Goldreich and Sridhar (\citeyear{GS95}, hereinafter GS95) proposed a theory on strong incompressible MHD turbulence, which predicts
Kolmogorov spectra
 \begin{equation}
    E(k) \propto k^{-5/3},
 \end{equation}
 for both velocity and magnetic field, and scale-dependent anisotropy
 \begin{equation}
    l_\| \propto l_\bot^{2/3},
 \end{equation}
where $l_\|$ and $l_\bot$ denote parallel and perpendicular sizes of an eddy, respectively.
Later, numerical simulations confirmed the GS95 scaling relations \citep{CV00ani,Maron01,ChoLV02}.
 
 In this paper we review strong Alfv\'enic MHD turbulence in various environments.
We consider not only the usual Alfv\'en wave but also its variants on small scales (i.e., whistler wave) and in the extreme relativistic limit (i.e., relativistic Alfv\'en wave).
 Except in Section \ref{sect:comp}, we assume incompressible fluids.
 In Section 2, we will briefly describe Kolmogorov theory for incompressible hydrodynamic turbulence.
 In Section 3, we consider strong incompressible Alfv\'enic MHD turbulence.
 In Section 4, we review weak incompressible MHD turbulence and show that, even in the case that
 large-scale turbulence is weak, strong MHD turbulence will ultimately  emerge on a small scale.
 In Section 5, we adopt the electron MHD (EMHD) model and describe strong magnetized turbulence on a scale below the proton gyro-scale, where whistler waves, i.e., variants of Alfv\'en waves in the EMHD regime, interact and generate turbulence. 
 In Sections 6 and 7, we briefly discuss strong relativistic Alfv\'enic turbulence and compressible MHD turbulence, respectively.
 In Section 8, we give conclusion.






\section{Hydrodynamic turbulence}
It may be useful to briefly review physical processes in incompressible hydrodynamic turbulence
before we proceed to MHD turbulence.
Detailed discussions on hydrodynamic turbulence can be found in \cite{Frisch95}.

Turbulence is present in many astrophysical fluids.
Why would we expect astrophysical fluids to be in turbulent state?
A fluid with viscosity $\nu$ becomes 
turbulent when the rate of viscous dissipation, which is $\sim \nu /L^2$ at the energy injection scale $L$,
 is much smaller than the energy transfer rate $\sim v_L/L$, where $v_L$  is the velocity dispersion at the scale $L$. The ratio of the two rates is the Reynolds number $Re = v_LL/\nu$. 
 In general, astrophysical fluids have $Re$'s much larger than critical values, which is usually 
 in the range from 10 to 100.  Therefore, we expect turbulence in astrophysical fluids.

\begin{figure}[ht]
	\centering
	\includegraphics[width=0.9\textwidth]{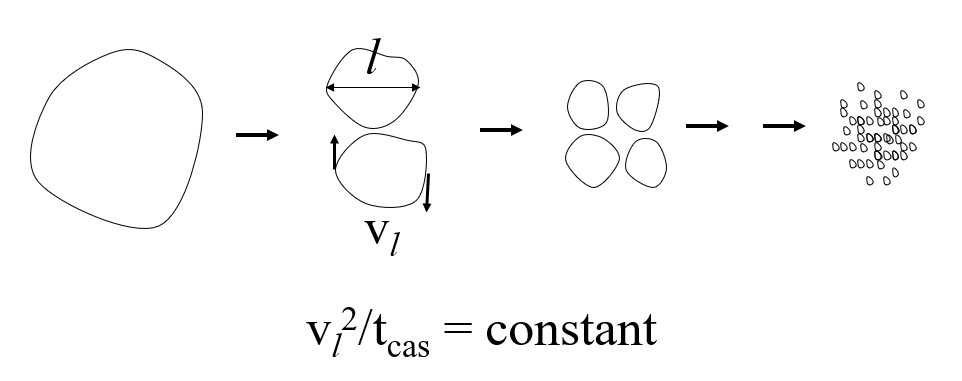}
	\caption{Energy cascade in hydrodynamic turbulence.}\label{fig:cascade}
\end{figure}

 There exists a well-known empirical theory, i.e., Kolmogorov theory  \citep{Kol41,Kol62}, for incompressible hydrodynamic turbulence.
 In fully developed turbulence, eddies of various sizes interact each other. 
 In Kolmogorov theory, kinetic energy contained in large-scale eddies cascades down to gradually smaller-scale eddies up to the viscosity scale (Figure \ref{fig:cascade}). In this picture, it is important to know that the energy cascade rate is scale-independent. Let us assume that turbulence is in a statistically stationary state and focus on a scale $l$. If energy cascade rate from a larger scale to $l$ is different from that from $l$ to a smaller scale, energy on the scale $l$ will either increase or decrease. In this case, we will not have a stationary state.
 Therefore, in order to have a stationary state, the energy cascade rate should be constant:
\begin{equation}
     v_l^2 /t_{cas} = \text{constant},  \label{eq:cas_rate}
\end{equation}
 where $v_l$ is velocity fluctuation on the scale $l$ and $t_{cas}$ is energy cascade time on the scale.
 Note that $v_l^2$ is proportional to kinetic energy, since density is constant in incompressible medium.
 
 It is possible to express $t_{cas}$ in terms of $l$ and $v_l$. A simple dimensional analysis gives
 \begin{equation}
    t_{cas} \sim l/v_l.   \label{eq:tcas}
 \end{equation}
 Physically speaking, $l/v_l$ can be viewed as a quantity proportional to the rotation period of an eddy of size $l$ and hence is called `eddy turnover time'.
 Note that $v_l$ is velocity difference between two points separated by $l$.
 If we consider an eddy of size $l$, the velocity difference between two points that lie on the boundary of the eddy and separated by $l$ is $\sim v_l$, which mean two points are moving away from each other at the speed of $v_l$ (see Figure \ref{fig:cascade}).
 In this picture,  the eddy will be completely distorted after $t\sim l/v_l$.
 Therefore, $l/v_l$ can also be viewed as distortion timescale of the eddy.
 
 If we substitute Equation (\ref{eq:tcas}) into Equation (\ref{eq:cas_rate}), we have
 \begin{equation}
   v_l^3/l =\text{constant},
 \end{equation}
or
 \begin{equation}
   v_l \propto l^{1/3}.
 \end{equation}
 This relation is valid from energy injection scale (a.k.a. outer scale of turbulence) to viscosity scale (a.k.a. inner scale of turbulence).
 
 If we perform Fourier transformation of velocity field $\mathbf{v}(\mathbf{r})$, we can obtain $\hat{\mathbf{v}}(\mathbf{k})$, where $\mathbf{k}$ is the wavevector.
 Then the (volume-integrated) energy spectrum is
 \begin{equation}
   E_v(k) = \int |\hat{\mathbf{v}}(\mathbf{k})|^2 d^3\mathbf{k},
 \end{equation}
 where the integration is done over a spherical shell of unit thickness and radius $k$.
 If the energy spectrum follows a power law, then we have the relation
 \begin{equation}
    kE_v(k) \sim v_l^2.   \label{eq:kEk}
 \end{equation}
 Since $v_l^2 \propto l^{2/3} \propto k^{-2/3}$, we have
 \begin{equation}
    E_v(k) \propto k^{-5/3},
 \end{equation}
which is called `Kolmogorov' spectrum.

\section{Strong Alfv\'enic MHD turbulence}    \label{sect:alf}
Let us consider MHD turbulence in a strongly magnetized incompressible medium.
As we will see in Section \ref{sect:cb_mhd}, 
in the presence of a strong magnetic field, two physically meaningful timescales exist: nonlinear hydrodynamic timescale and wave-crossing timescale.
In this section, we consider a special regime of turbulence called `strong' MHD turbulence, in which two timescales have very similar values. 
If turbulence is driven isotropically, we need $v_L \sim V_A$ in order to have strong turbulence near the driving scale, 
where $v_L$ is the velocity fluctuation on the driving scale $L$ and $V_A$ is the Alfv\'en speed.
Note that $V_A=B_0$ in the system of units we adopt (i.e., numerical units) and
 $V_A=B_0/\sqrt{4 \pi \rho}$ in usual units.
In the next section, we will show that `strong' MHD turbulence will appear below a scale smaller than $L$ 
even in the case of  $v_L < V_A$ in isotropically driven turbulence. 
Therefore, strong MHD turbulence may be regarded as a common type of MHD turbulence.

\subsection{Interaction of Alfv\'en wave packets}
It is well known that magnetic field lines have tension. 
Therefore, we may regard a magnetic field line as an elastic string.
The restoring force for Alfv\'en wave is magnetic tension.
If we properly perturb an elastic string, we can make the perturbation move along the string.
Likewise, if we properly perturb magnetic field lines, we can generate an Alfv\'en wave packet that oscillates due to magnetic tension and we can make it move along the field lines.
Note that we can make the Alfv\'en wave packet move either parallel or anti-parallel to the mean magnetic field.

The speed of Alfv\'en wave packet is $B_0$ in numerical units or $B_0/\sqrt{4 \pi \rho}$ in usual units.
Since all the Alfv\'en wave packets move at the same speed along the magnetic field, they do not interact. Therefore, Alfv\'en wave packets moving in one direction do not change their shapes in time, which means we can not have turbulence out of wave packets moving in one direction.

Mathematically, we can describe motion of Alfv\'en wave packets moving in one direction as follows.
The MHD equations in incompressible medium are 
\begin{eqnarray}
 \frac{\partial \mathbf{v}}{\partial t} = -\mathbf{v} \cdot \nabla \mathbf{v} + ( B_0 \hat{\mathbf{z}}+\mathbf{b} ) \cdot \nabla \mathbf{b} - \nabla P,\\
 \frac{\partial \mathbf{b}}{\partial t} = -\mathbf{v} \cdot \nabla \mathbf{b}+\mathbf{b} \cdot \nabla \mathbf{v},
\end{eqnarray}
where, $\mathbf{v}$ is velocity, $B_0$ the strength of the mean magnetic field, $\hat{\mathbf{z}}$ the unit vector along z-direction,  $\mathbf{b}$ fluctuating magnetic field, and  $P$ the total pressure. We use numerical units for magnetic field, which means $\mathbf{b}$ actually represents $\mathbf{b}/\sqrt{4 \pi \rho}$. For simplicity, we ignore the viscosity and the magnetic dissipation terms.
If we rewrite the equations in terms of Elss\"asser variables, $\mathbf{Z}^+ \equiv B_0 \hat{\mathbf{z}}+\mathbf{v}+\mathbf{b}$ and
$\mathbf{Z}^- \equiv B_0 \hat{\mathbf{z}}+\mathbf{v}-\mathbf{b}$, we have
\begin{eqnarray}
 \frac{\partial \mathbf{Z}^+}{\partial t} -B_0 \frac{\partial \mathbf{Z}^+}{\partial z} = - \mathbf{Z}^- \cdot \nabla \mathbf{Z}^+ - \nabla P,  \label{eq:zplus} \\
  \frac{\partial \mathbf{Z}^-}{\partial t} +B_0 \frac{\partial \mathbf{Z}^-}{\partial z} = - \mathbf{Z}^+ \cdot \nabla \mathbf{Z}^- - \nabla P.
  \label{eq:zminus}
\end{eqnarray}
If $\mathbf{Z}^-$ is zero, the nonlinear term in  Equation (\ref{eq:zplus}) vanishes and we have
\begin{equation}
 \frac{\partial \mathbf{Z}^+}{\partial t} -B_0 \frac{\partial \mathbf{Z}^+}{\partial z} = - \nabla P. \\
\end{equation}
If we ignore the pressure gradient term\footnote{
   The variable $\mathbf{Z}^+$ is incompressible, i.e., $\nabla \cdot \mathbf{Z}^+ =0$, while the pressure gradient term is purely compressible.
   Therefore, generally speaking the pressure gradient term does not directly affect time evolution of $\mathbf{Z}^+$ in incompressible medium.
 It is also possible to show that $\nabla P$ is negligibly small in an Alfv\'enic perturbation with a small amplitude.
},
this becomes a wave equation for $\mathbf{Z}^+$.
The wave speed is constant (i.e., $B_0$), which means all waves have the same speed.
If we seek a solution of the form $\mathbf{Z}(k_z z+\omega t)$, we have $\omega=k_z B_0$, or
\begin{equation}
      \omega =k_z V_A,    \label{eq:omega_alf}
\end{equation}
which means Alfv\'en waves are not dispersive and therefore Alfv\'en waves moving in one direction do not interact each other.

\begin{figure}[ht]
	\centering
	\includegraphics[width=0.9\textwidth]{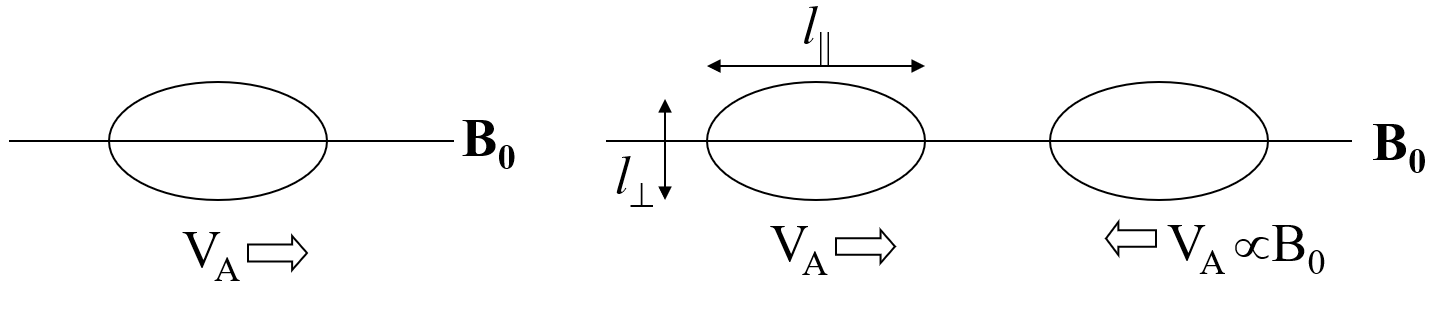}
	\caption{Alfv\'en wave packets. (Left) An Alfv\'en wave packet moves at the Alfv\'en speed $V_A=B_0/\sqrt{ 4\pi \rho}$. When we use a system of units in which $4 \pi$ does not appear and $\rho=1$, it follows that $ 4 \pi \rho =1$ and hence $V_A=B_0$. Since Alfv\'en waves are non-dispersive, an Alfv\'en wave packet does not change its shape during the propagation in an incompressible fluid.
(Right) When two opposite-traveling Alfv\'en wave packets collide, non-linear interactions happen.
    }\label{fig:wave}
\end{figure}

Now, let us consider two opposite-traveling wave packets (see Figure \ref{fig:wave}).
When they collide, they interact. Then, how can they interact?
If we look at the time evolution of $\mathbf{Z}^+$ (see Equation (\ref{eq:zplus})), 
the nonlinear interaction term $ \mathbf{Z}^- \cdot \nabla \mathbf{Z}^+$ states that shearing motion provided by $ \mathbf{Z}^-$ distorts $ \mathbf{Z}^+$. 
Through the nonlinear process, smaller structures can be generated. 
If we drive turbulence on scale $L$, driving force generates opposite-traveling Alfv\'enic wave packets on scale $L$ and collisions between them create smaller wave packets.
Then, collisions between newly created smaller wave packets create even smaller wave packets.
This `cascade' process continues up to the dissipation scale and, this way, we have Alfv\'enic turbulence cascade.

\subsection{Critical balance}  \label{sect:cb_mhd}
In the previous subsection, we showed that collision of opposite-traveling wave packets (or `eddies') is essential for Alfv\'enic turbulence.
Based on the concept of critical balance, GS95 \citep{GS95} developed a theory on strong Alfv\'enic MHD turbulence.
In this and next subsections, we briefly introduce critical balance and scaling relations by GS95.

\begin{figure}[ht]
	\centering
	\includegraphics[width=0.9\textwidth]{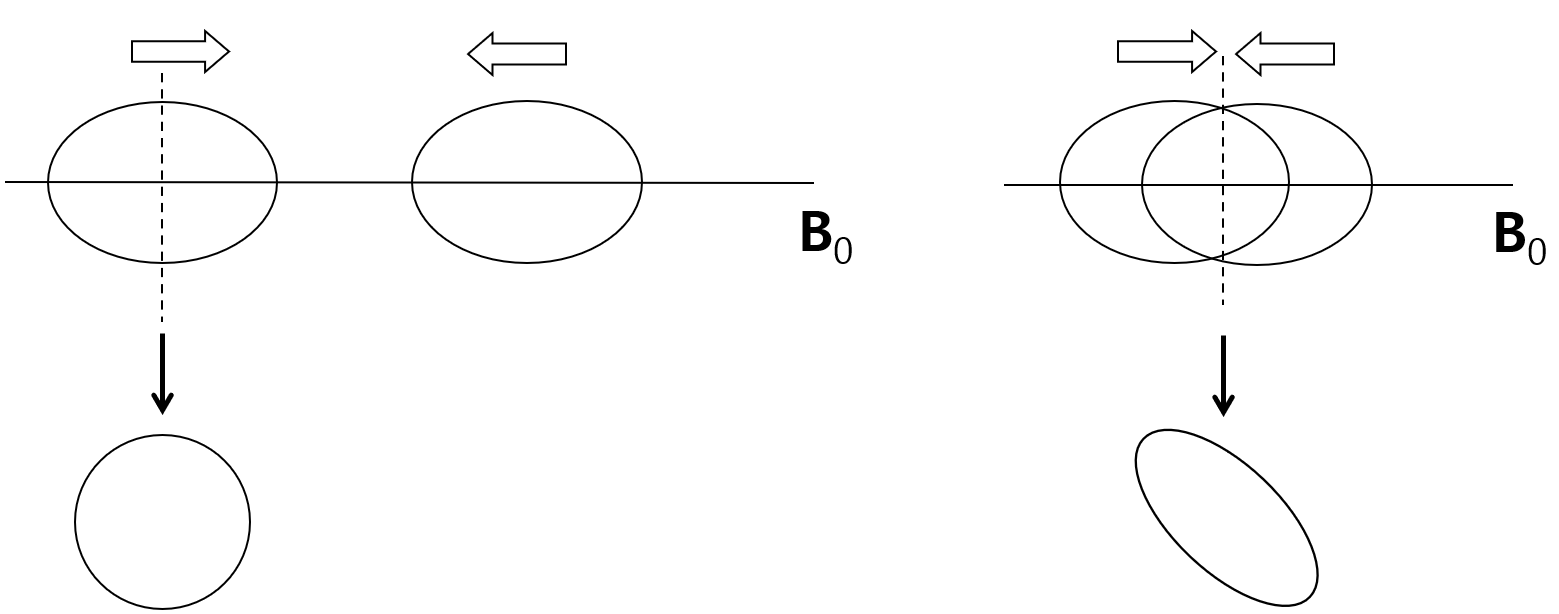}
	\caption{Collision of wave packets. Shapes shown below the arrows are cross-sectional shapes. (Left) Before the collision, we assume perpendicular shapes of the eddies are isotropic. (Right) During the collision, an eddy is distorted by shearing motion of the other eddy.}\label{fig:coll}
\end{figure}

 Let us consider two opposite-traveling wave packets with similar sizes (see Figure \ref{fig:wave}). Let us assume that their sizes in the directions parallel and perpendicular to $\mathbf{B}_0$ are $l_\|$ and $l_\bot$, respectively.
We assume the cross-sectional shapes of the eddies are almost circular (i.e., isotropic) before the collision (see the left panel of Figure \ref{fig:coll}).
During the collision, an eddy is distorted by shearing motion of the other eddy. 
Here, we only consider distortion in the perpendicular direction.
If velocity difference between two points, whose perpendicular separation is $l_\bot$, is $v_l$, then the distortion timescale is 
\begin{equation}
    t_{eddy} \sim l_\bot/v_l,   \label{eq:eddy}
\end{equation}
which is none other than the eddy turnover time of hydrodynamic turbulence.
Note that, when an eddy is completely distorted, it will become different scale (i.e., smaller scale) eddies through nonlinear interactions, which means energy cascade is completed from scale $l_\bot$ to a smaller scale.
On the other hand, the duration of collision is equal to the parallel size divided by the wave speed:
\begin{equation}
   t_{wave} \sim l_{\|}/V_A.    \label{eq:wave}   
\end{equation}

If $t_{eddy} \sim t_{wave}$, an eddy becomes completely distorted after one collision, which means that one collision is enough
to complete energy cascade. 
The condition  $t_{eddy} \sim t_{wave}$ is called `critical balance' and `critically balanced' turbulence  is called `strong' MHD turbulence.
From Equations (\ref{eq:eddy}) and (\ref{eq:wave}), the condition for strong MHD turbulence becomes
\begin{equation}
     \frac{ l_\bot }{ v_l } \sim \frac{  l_\| }{V_A}.     \label{eq:critical}
\end{equation}
If turbulence driving is isotropic, we expect $L_\| \sim L_\bot$ at the driving scale $L$.
In this case, the condition for strong MHD turbulence at the driving scale is
\begin{equation}
    V_A  \sim v_L,     \label{eq:critical_iso}
\end{equation}
which means that the Alfv\'en Mach number is unity:
\begin{equation}
   M_A \equiv \frac{v_L}{V_A} \sim 1.
\end{equation}

The condition for critical balance in Equation (\ref{eq:critical}) or (\ref{eq:critical_iso}) looks very special and hard to fulfill.
However, in next section, we will show that even in the case of  $t_{eddy} > t_{wave}$
on the driving scale, we will ultimately have the condition met on smaller scales.

\subsection{Scaling relations in strong Alfv\'enic turbulence}  \label{sect:scale_mhd}
As in hydrodynamic turbulence, we can assume constancy of energy cascade rate.
By combining constancy of energy cascade rate and critical balance, we can derive energy spectrum and anisotropy of strong Alfv\'enic turbulence:
\begin{itemize}
		\setlength{\itemindent}{.5in}
	\item Constancy of energy cascade rate:
	 \begin{equation}
          v_l^2/t_{cas}=\text{constant},  \label{eq:cas_rate_alf}
     \end{equation}  \\
   \item Critical balance:
      \begin{equation}
  \frac{ l_\bot }{ v_l } \sim \frac{  l_\| }{V_A}.  \label{eq:cbal2}
      \end{equation}
\end{itemize}
In Equation (\ref{eq:cas_rate_alf}), $v_l^2$ is proportional to kinetic energy density on scale $l$. Since $v_l \sim b_l$ in Alfv\'en wave packets, either $v_l^2$ or $b_l^2$ can be used in the equation.

There are two physically meaningful timescales in MHD: hydrodynamic timescale 
($t_{eddy} \sim l_\bot/v_l$) and wave-crossing timescale 
($t_{wave} \sim l_\| /V_A$). Since two timescales are virtually same (see Equation (\ref{eq:cbal2})) in
strong turbulence, we can use either of them for $t_{cas}$.
If we use $ l_\bot/v_l$ for $t_{cas}$, Equation (\ref{eq:cas_rate_alf}) becomes
\begin{equation}
  v_l^3/l_\bot =\text{constant}.   \label{eq:b_l}
\end{equation}
As in hydrodynamic turbulence, the kinetic energy spectrum corresponding to Equation (\ref{eq:b_l}) is
\begin{equation}
  E_v(k) \propto k^{-5/3}.
\end{equation}
Since $v_l \sim b_l$ in Alfv\'en wave packets, we also have a Kolmogorov spectrum for magnetic field:
\begin{equation}
  E_b(k) \sim E_v(k) \propto k^{-5/3}.
\end{equation}

If we substitute Equation (\ref{eq:b_l}) into Equation (\ref{eq:cbal2}), we have
\begin{equation}
   l_\| \propto l_\bot^{2/3}.   \label{eq:ani}
\end{equation}
Note that $l_\|$ and $l_\bot$ denote parallel and perpendicular sizes of an eddy, respectively.
Therefore, Equation (\ref{eq:ani}) tells us anisotropy of eddy shape:
smaller eddies are more elongated along the (local) mean field direction.

In summary, we use constancy of energy cascade rate and critical balance to derive scaling relations of strong Alfv\'enic turbulence. The result shows that strong Alfv\'enic turbulence has Kolmogorov spectra for both velocity and magnetic field and scale-dependent anisotropy of eddy shape.

\subsection{Numerical tests and observations}
Using numerical simulations, \cite{CV00ani} first tested the theory of strong MHD turbulence.
We numerically solved the
incompressible MHD equations in a periodic box and calculated spectrum and anisotropy.
We used isotropic large-scale driving to generate strong MHD turbulence.

\begin{figure}[ht]
  \centering
\begin{minipage}{.4\textwidth}
	\centering
	\includegraphics[width=.94\linewidth]{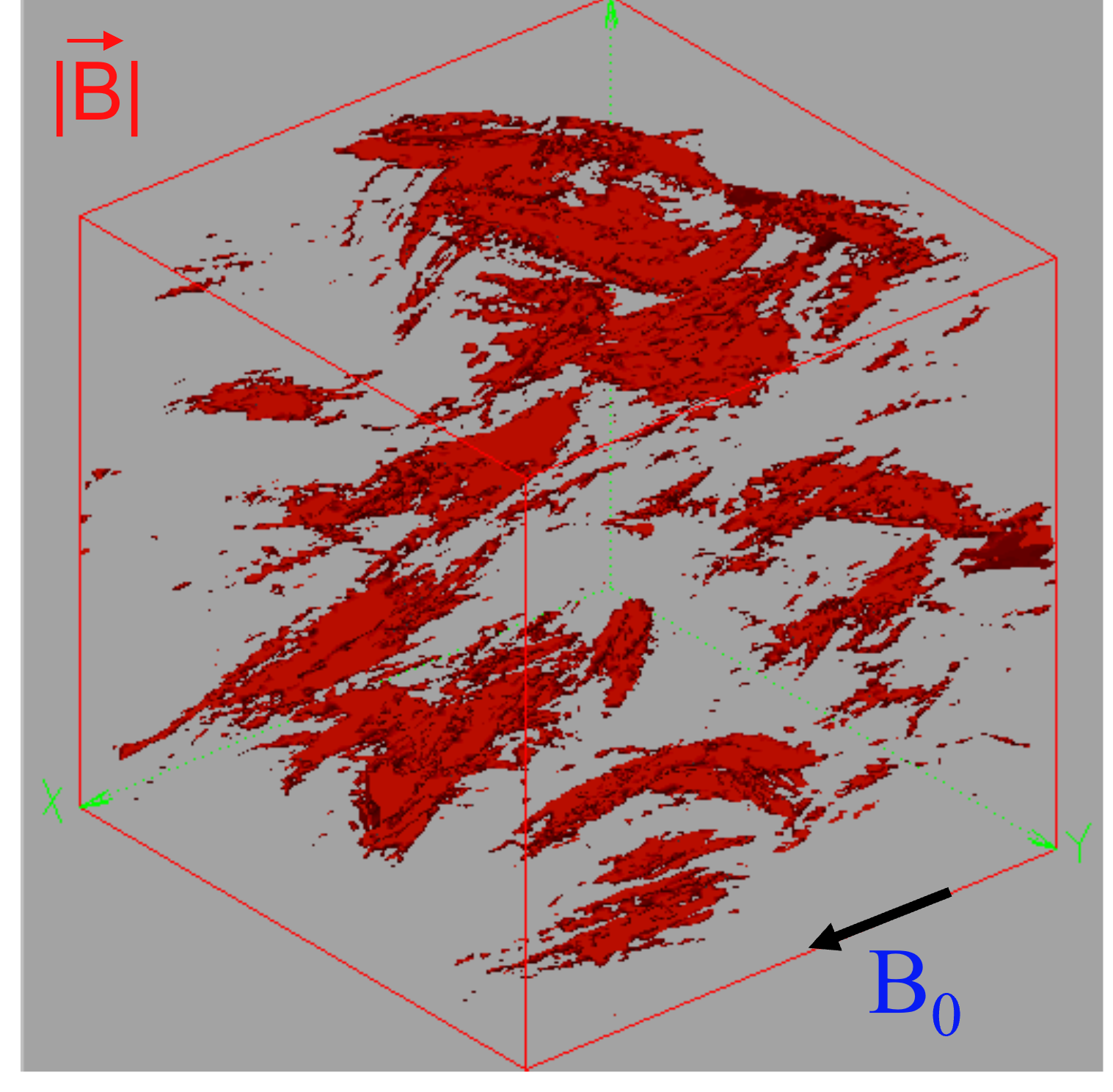}
		   \vspace{3mm}
\end{minipage}%
\begin{minipage}{.6\textwidth}
	\centering
	\includegraphics[width=.9\linewidth]{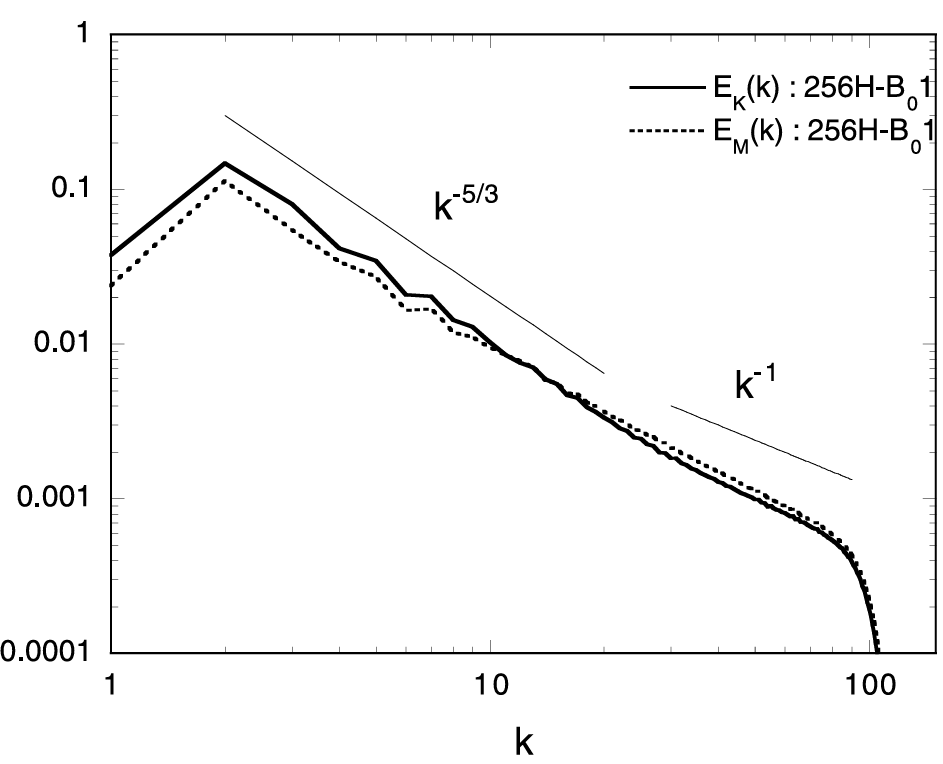}
\end{minipage}
\caption{(Left) Snapshot of magnetic field strength. The black arrow represents the direction of the mean magnetic field.
      (Right) Kinetic and magnetic spectra. They are compatible with Kolmogorov spectrum and hence
     agree with the GS95 prediction. From \cite{CV00ani}.
}%
\label{fig:sp_cv2000}%
\end{figure}

Figure \ref{fig:sp_cv2000} shows results from a simulation in \cite{CV00ani}.
The left panel of Figure \ref{fig:sp_cv2000} visualizes three-dimensional distribution of
$|\mathbf{B}|$.
The right panel of the figure shows kinetic and magnetic spectra.
Although the magnetic spectrum (dotted line) is slightly shallower than the kinetic one (solid line), the overall spectra are roughly  consistent with Kolmogorov spectrum.

\begin{figure}[ht]
  \centering
\begin{minipage}{.38\textwidth}
	\centering
	\includegraphics[width=.97\linewidth]{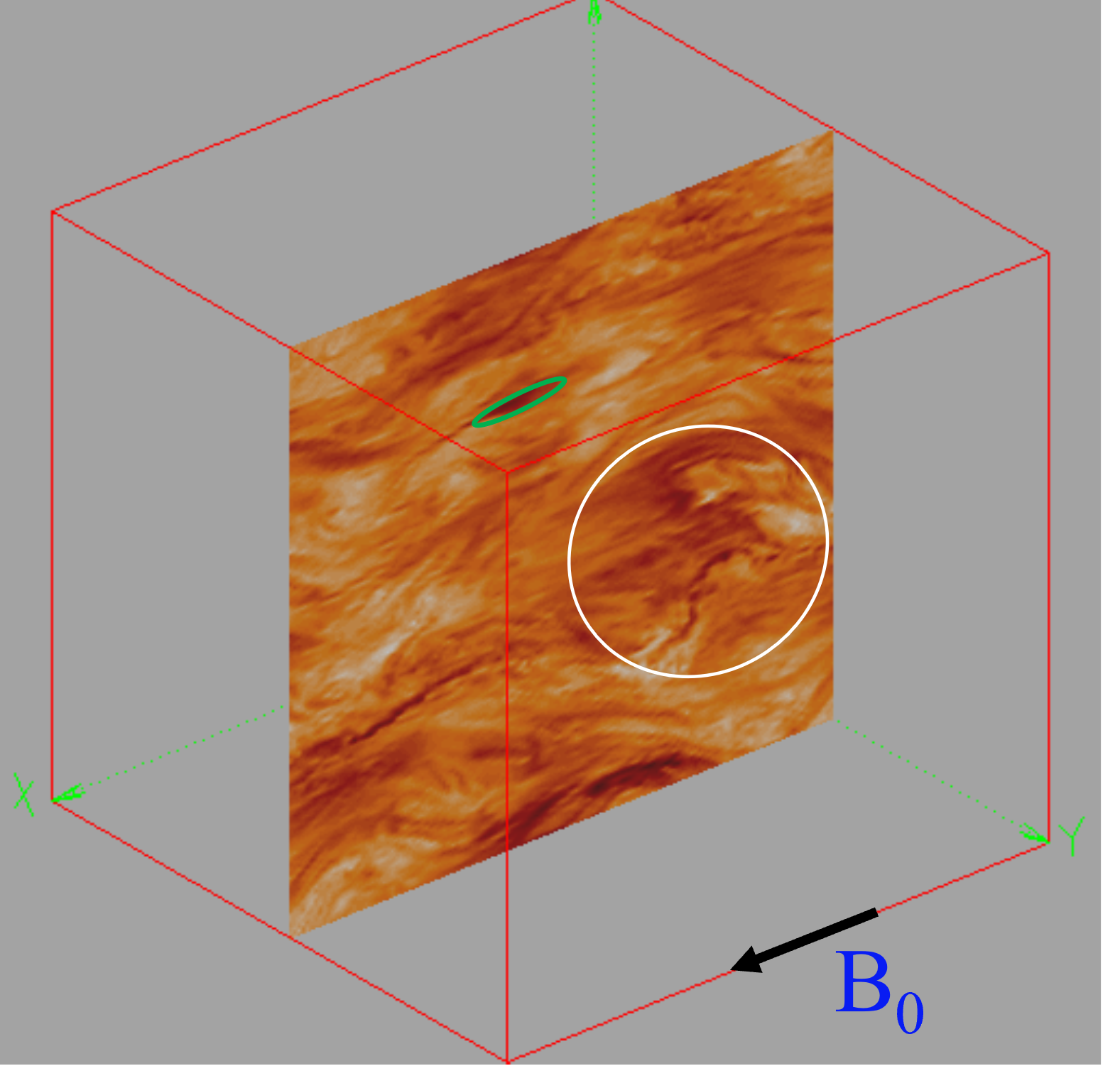}
		   \vspace{4mm}
\end{minipage}%
\begin{minipage}{.62\textwidth}
	\centering
	\includegraphics[width=.95\linewidth]{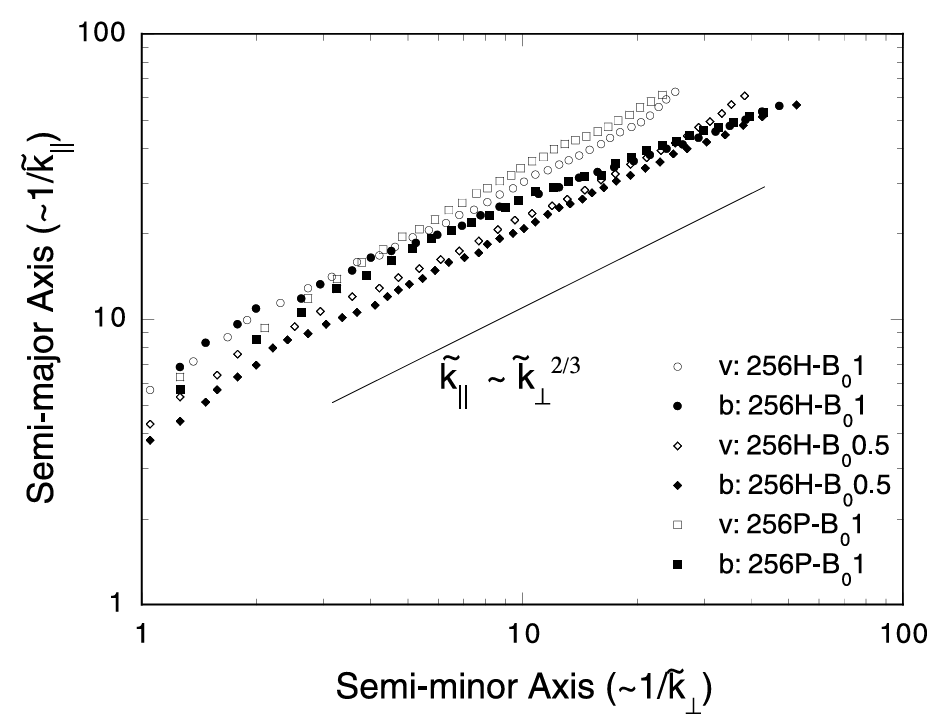}
\end{minipage}
\caption{(Left) Illustration of scale-dependent anisotropy. The black arrow represents the direction of the mean magnetic field.
 (Right) Measured anisotropy of eddy shapes. The parallel and perpendicular sizes of eddies, $l_\|$ and $l_\bot$ respectively, follow the relation $l_\| \propto l_\bot^{2/3}$, which agrees with the GS95 prediction. From \cite{CV00ani}.
}%
\label{fig:ani_cv2000}%
\end{figure}

Figure \ref{fig:ani_cv2000} shows anisotropy of strong MHD turbulence.
The left panel of Figure \ref{fig:ani_cv2000} demonstrates scale-dependent anisotropy.
Large-scale eddies (see, for example, the white ellipse) are more or less isotropic.
However, small-scale eddies  (see, for example, the green ellipse) are highly elongated.
\cite{CV00ani} devised a method to quantitatively measure the parallel and perpendicular sizes of eddies. 
According to GS95, the parallel and perpendicular sizes of eddies follow the relation $l_\| \propto l_\bot^{2/3}$.
The right panel of Figure \ref{fig:ani_cv2000} shows that the relation holds true.
The horizontal and vertical axes correspond to $l_\bot$ and $l_\|$, respectively.

Measurements of solar wind magnetic spectrum have a long history
(for review, see, e.g., \citealt{Bruno13}).
Although there are some uncertainties, the results are mostly compatible with the GS95 prediction.
Scale-dependent anisotropy has been measured only recently and
the results are consistent with the GS95 anisotropy (see, for example, \citealt{Horbu08, Podes09,Wicks10}; see also a review by \citealt{Chen16review}).

\subsection{Further developments in strong Alfv\'enic turbulence research}
In this section, we only focus on simple and basic physics of strong MHD turbulence.
However, there are many complicated aspects of strong MHD turbulence.
For example, when we considered collision of opposite-traveling wave packets, we tacitly assumed that they have similar amplitudes (i.e., similar values for $|\mathbf{b}_l|$). However, in reality, they can have different amplitudes. This kind of turbulence is called `imbalanced' turbulence
\citep{Lithw07,BeresL08,BeresL09,Chand08,PerezB09,PodesB10,Perez12,Mason12,ChoL14}.
There are also claims that alignment between variables (e.g., Elss\"asser variables $\mathbf{Z}^+$ and
$\mathbf{Z}^-$) can modify energy spectrum of strong MHD turbulence
\citep{Boldy05,Boldy06,BeresL06,Beatt25}.
There are also other issues related to strong MHD turbulence
(see reviews by \cite{Beres19review} and \cite{Schek22review} for further details).

\section{MHD turbulence  with a very strong mean magnetic field}  \label{sect:weak}
In the previous section, we introduced strong Alfv\'enic MHD turbulence, which is based on critical balance.
In case of isotropic driving, the condition for critical balance is $V_A \sim v_L$.
However, if the mean magnetic field is very strong, it is possible to have $V_A > v_L$ for isotropic driving, or $V_A L_\bot / v_L L_\| > 1$ for general cases.
This type of turbulence is called `weak' MHD turbulence.
In case of isotropic driving, the Alfv\'en Mach number ($M_A\equiv v_{rms}/V_A$) is less than unity in weak turbulence.
Therefore, weak MHD turbulence virtually coincides with sub-Alfv\'enic turbulence if driving is isotropic.
In this section, we will show that critical balance is satisfied below the scale $\sim LM_A^2$,  even if $V_A \gg v_L$.

\cite{Zakha92} provided general discussions on weak turbulence.
 \cite{GS97} and \cite{NgB97}
 derived the spectrum of 
 weak turbulence using scaling arguments.
Galtier (\citeyear{Galti00}) first developed kinetic theory for weak turbulence.

\subsection{Cascade time in weak turbulence}
In weak turbulence, $t_{eddy} > t_{wave}$.
Since the time necessary for complete distortion (i.e., $t_{eddy}$) is longer than duration of collision (i.e., $t_{wave}$), 
an eddy cannot be completely distorted after one collision, which means that more than one collisions are necessary
for complete distortion of the eddy. 
Therefore, more than one collision is necessary
to complete energy cascade. 

Then how many collisions are necessary?
The magnetic induction equation reads
\begin{equation}
   \frac{\partial \mathbf{B}}{\partial t} =\nabla \times ( \mathbf{v} \times \mathbf{B} ) + \eta \nabla^2 \mathbf{B},
\end{equation}
where $\mathbf{B}=\mathbf{B}_0 + \mathbf{b}$ and $\eta$ is magnetic diffusivity.
From this, we can write
\begin{equation}
  \frac{db_l}{dt} \sim \frac{b_l v_l}{l_\bot},  \label{eq:dbdt}
\end{equation}
where we assume derivative in the direction perpendicular to the (local) mean magnetic field is larger than the one in the parallel direction due to anisotropy of turbulence in the presence of strong mean magnetic field.
Here we also assume 
locality of nonlinear interactions: i.e., time evolution of an eddy of size $l$ is determined by interactions with similar-size eddies and thus the magnitude of
the nonlinear term on scale $l$ is $|\nabla \times (\mathbf{v} \times \mathbf{b})| \sim v_l b_l/l_\bot$.
Since $v_l \sim b_l$, we can rewrite (\ref{eq:dbdt}) as follows:
\begin{equation}
  \frac{dv_l}{dt} \sim \frac{v_l^2}{l_\bot}.  \label{eq:dvdt}
\end{equation}

From Equation (\ref{eq:dvdt}), we have
\begin{equation}
  \frac{d\mathcal{E}_l}{dt} \sim \frac{v_l^3}{l_\bot},  \label{eq:db2dt}
\end{equation}
where $\mathcal{E}_l \sim v_l^2$ is energy density on scale $l$.
The change in energy during one collision is 
\begin{equation}
 | \Delta \mathcal{E}_l | \sim (d\mathcal{E}_l/dt) \Delta t \sim (v_l^3/l_\bot )t_{wave} \sim  (v_l^3/l_\bot )(l_\| /V_A) ,
\end{equation}
and, therefore, we have
\begin{equation}
   \frac{ | \Delta \mathcal{E}_l | }{ \mathcal{E}_l } 
     \sim \frac{ v_l l_\| }{ V_A l_\bot } = \frac{ t_{wave} }{ t_{eddy} } \equiv \chi < 1. \label{eq:dee}
\end{equation}

Equation (\ref{eq:dee}) confirms that many collisions are necessary to complete energy cascade in weak Alfv\'enic turbulence.
If each collision contributes to energy change randomly, then the number of collisions required for energy cascade will be
\begin{equation}
  N \sim (1/\chi)^2.
\end{equation}
Therefore, the cascade time becomes
\begin{equation}
  t_{cas} \sim N t_{wave} \sim (t_{eddy}/t_{wave})^2 t_{wave} =(t_{eddy}/ t_{wave}) t_{eddy} = (V_A l_\bot^2 / v_l^2 l_\| ).   \label{eq:tcas_weak}
\end{equation} 

\subsection{Anisotropy of weak MHD turbulence}
Since $t_{eddy} > t_{wave}$, wave-like motions and wave-wave interactions are important in weak turbulence.
Many researchers have argued that wave-wave interactions result in strong anisotropy (see, for example, \citealt{Sheba83, SG94, Montg95}).  Mathematically, anisotropy manifests itself in
 the resonant conditions for 3-wave interactions:
 \begin{eqnarray}
 \mathbf{k}_1 +\mathbf{k}_2 = \mathbf{k}_3,   \label{eq:kvec3}    \\
 \omega_1 +\omega_2 = \omega_3,  \label{eq:omega3}
 \end{eqnarray}
 where $\mathbf{k}$'s are wavevectors and $\omega$'s are wave frequencies.
 If we rewrite Equation (\ref{eq:kvec3}) only for the parallel direction and 
               Equation (\ref{eq:omega3}) using 
$\omega = V_A | k_\| | $ (see Alfv\'en-wave dispersion relation in Equation (\ref{eq:omega_alf})), we have
\begin{eqnarray}
   k_{\|,1}+k_{\|,2} = k_{\|,3},  \label{eq:k3} \\
  |k_{\|,1}|+|k_{\|,2}| = |k_{\|,3}|,   \label{eq:lkl3}
\end{eqnarray}
where $k_\|$'s are parallel wavenumbers.

Since only opposite-traveling wave packets interact, $k_{\|,1}$ and $ k_{\|,2}$ must have opposite signs. 
Suppose that $k_{\|,3}$ is positive.
In this case, there two solutions for Equations (\ref{eq:k3}) and (\ref{eq:lkl3}): 
     either $k_{\|,1}=0$ and $k_{\|,2}=k_{\|,3}$ or $k_{\|,2}=0$ and $k_{\|,1}=k_{\|,3}$.
In the case of $k_{\|,1}=0$ and $k_{\|,2}=k_{\|,3}$, energy in wave 2 is transferred to wave 3 by the help of wave 1.
Since $k_\|$ of `donor' is the same as that of `recipient', the parallel wavenumber ($k_\|$) does not change during the cascade:
\begin{equation}
   k_\| =\text{constant},  \label{eq:kconst}
\end{equation}
 which means that the parallel size
of an eddy does not change during the cascade. Only the perpendicular size reduces during the cascade.

\subsection{Scaling relation of weak MHD turbulence}
By combining constancy of energy cascade rate, the expression for $t_{cas}$ in Equation (\ref{eq:tcas_weak}), and anisotropy in Equation (\ref{eq:kconst}), we can derive energy spectrum and anisotropy of weak Alfv\'enic turbulence:
\begin{itemize}
		\setlength{\itemindent}{.5in}
	\item Constancy of energy cascade rate:
	 \begin{equation}
          v_l^2/t_{cas}=\text{constant},  \label{eq:cas_rate_weak}
     \end{equation}  
   \item Expression for $t_{cas}$:
      \begin{equation}
          t_{cas} \sim \left( \frac{ t_{eddy} }{ t_{wave} } \right) t_{eddy} = \frac{ V_A  l_\bot^2 }{v_l^2 l_\|}.  \label{eq:tcas_weak2}
      \end{equation}
   \item Anisotropy:
      \begin{equation}
            k_\| =\text{constant}.  \label{eq:kconst2}
      \end{equation}
\end{itemize}
If we insert Equation (\ref{eq:tcas_weak2}) into Equation (\ref{eq:cas_rate_weak}), we have
\begin{equation}
    v_l^2 (v_l^2 l_\| / V_A l_\bot^2 ) = \text{constant, or }  v_l^2 \propto  l_\bot,   \label{eq:bsq_weak}
\end{equation}
where we use the fact that $k_\|$ is constant (Equation (\ref{eq:kconst2})). 
Since $kE_v(k) \sim v_l^2$ (see Equation (\ref{eq:kEk})), we have the following energy spectrum:
\begin{equation}
  E_v(k) \propto k_\bot^{-2}.
\end{equation}
Since $v_l \sim b_l$ for Alfv\'enic perturbations, we also have a $k_\bot^{-2}$ spectrum for magnetic field.

\subsection{Transition form weak to strong turbulence on a small scale}
\begin{figure}[ht]
	\centering
	\includegraphics[width=0.8\textwidth]{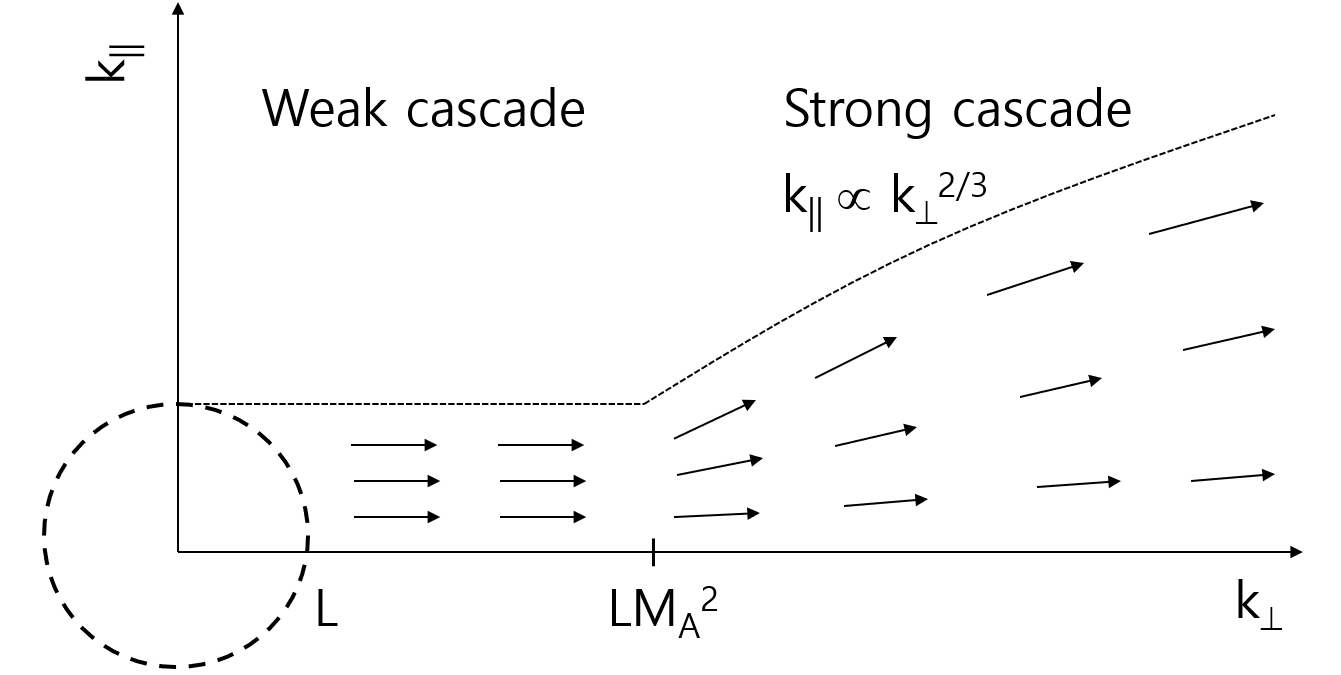}
	\caption{Transition from weak to strong cascade on the scale $\sim LM_A^2$, where $M_A=v_L/B_0$ is the Alfv\'en Mach number.}\label{fig:transition}
\end{figure}

In weak turbulence, $\chi < 1$ on the driving scale.
From Equation (\ref{eq:bsq_weak}), we have
\begin{equation}
  v_l = v_L (l_\bot/L)^{1/2},
\end{equation}
where $v_L \sim v_{rms}$ is the velocity on the driving scale $L$.
Using this,
we can write $\chi$ on scale $l_\bot$:
\begin{equation}
  \chi =  \frac{v_l l_\| }{ V_A l_\bot } = \frac{ v_L L_\| }{ V_A l_\bot^{1/2} L^{1/2} } \sim \frac{ v_L }{ V_A } \left( \frac{ L }{ l_\bot } \right)^{1/2},
\end{equation}
where we use $l_\| = L_\| \sim L$.
This equation tells us that $\chi$ increases as $l_\bot$ decreases and 
$\chi$ becomes unity on the scale
\begin{equation}
   l_\bot \sim (v_L/V_A)^2 L = M_A^2 L.
\end{equation}
Therefore, transition from weak to strong turbulence happens on this scale.
Below the transition scale, critical balance is satisfied and we can use the scaling relations for strong
MHD turbulence.
In Figure \ref{fig:transition}, we illustrate this transition in Fourier space. Transition from weak to strong turbulence is observed in direct numerical simulations \citep{Meyra16} and also in space plasmas \citep{Zhao24}.


\section{Small-scale MHD Turbulence}    \label{sect:small}
We can use MHD to describe large-scale motions.
However, on sufficiently small scales, MHD approximation will ultimately fail and
 we cannot apply the theory of MHD turbulence to the small scales.
In the solar wind, for example, the slope of magnetic energy spectrum changes near the proton gyro-scale.
The magnetic energy spectrum is consistent with Kolmogorov one above the gyro-scale, while it is much steeper than Kolmogorov one below the gyro-scale
(see illustration in the left panel of Figure \ref{fig:emhd}).
The scales below the MHD scale are sometimes called `dispersion range' since waves in the range are dispersive.
The power index of the energy spectrum in the dispersion range usually varies between $-2$ and $-4$
(see, for example, \citealt{Leamo98,Leamo99,Smith06}).
Understanding physics of the dispersion range is of great importance in space physics \citep{Dmitr06,Howes08a,Matth08,Howes08b,Saito08, Schek09}.

In this section, we use Electron MHD (EMHD) to understand physics of small-scale turbulence.
EMHD is a very simple fluid-like model to describe small-scale physics \citep{Kingsep1990review}.
As early as 1990s, researchers found a steep energy spectrum ($E(k) \propto k^{-7/3}$) for EMHD turbulence \citep{Biska96, Biska99, NgB03}.
There are also studies on anisotropic nature of (two-dimensional) EMHD turbulence \citep{Dastg03, NgB03}.
In 2004, \citeauthor{ChoL04emhd} first derived scale-dependent anisotropy of three-dimensional strong EMHD turbulence based on critical balance.
In this section, we only consider strong EMHD turbulence.
Discussions on weak EMHD turbulence can be found in \cite{GaltiB03}, \cite{Galti05} and \cite{Galti06}.

\subsection{Electron MHD model}

\begin{figure}[ht]
	\centering
	\includegraphics[width=0.9\textwidth]{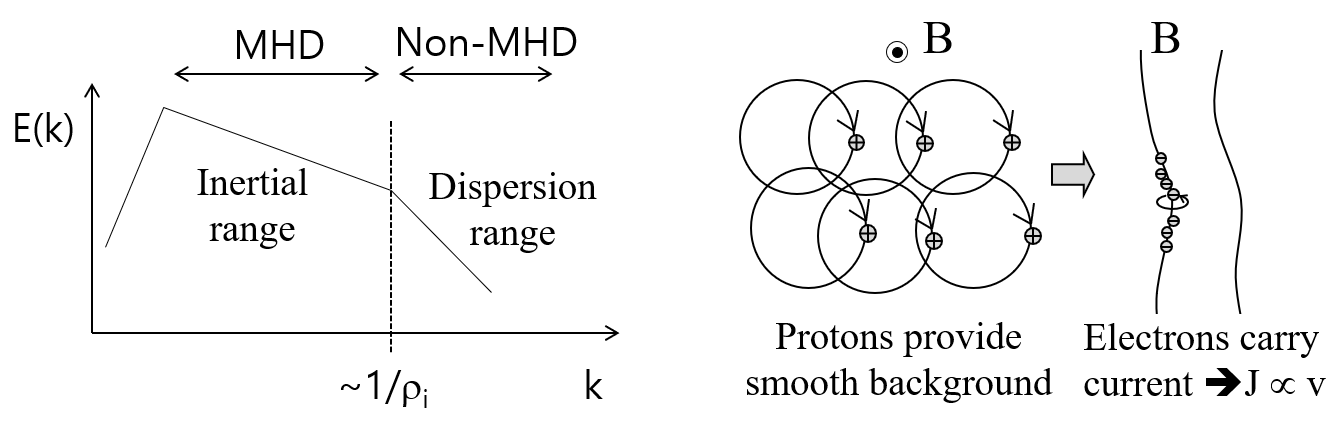}
	\caption{(Left) Change of energy spectrum near the proton gyro-scale $\rho_i$. (Right) Electron MHD model. Protons provide smooth charge background and only electrons carry current.}\label{fig:mhd_emhd}
\end{figure}

EMHD assumes that protons provide smooth charge background and only electrons carry current on scales below the proton gyro-scale 
(see the right panel of Figure \ref{fig:mhd_emhd}).
Therefore, since electron motions generate current, current density is proportional to electron velocity:
\begin{equation}
  \mathbf{J} = -ne \mathbf{v},  \label{eq:J_emhd}
 \end{equation}
 where $n$ is electron number density, $e$ is electric charge, and $\mathbf{v}$ is electron velocity.
 Note that electron velocity here does not mean velocity of individual electron, but bulk velocity of electron `fluid'.
 From Equation (\ref{eq:J_emhd}), we have
 \begin{equation}
  \mathbf{v}= -\frac{ \mathbf{J} }{ne} = -\frac{ c }{ 4 \pi ne} \nabla \times \mathbf{B} .  \label{eq:v_emhd}
\end{equation}

To derive the EMHD equation, let us start with magnetic induction equation
\begin{equation}
   \frac{ \partial \mathbf{B} }{ \partial t} = \nabla \times ( \mathbf{v} \times \mathbf{B} ) + \eta \nabla^2 \mathbf{B}.  \label{eq:induction}
\end{equation}
If we substitute Equation (\ref{eq:v_emhd}) into Equation (\ref{eq:induction}), we have
\begin{equation}
    \frac{ \partial \mathbf{B} }{ \partial t} = -\frac{ c }{ 4 \pi ne} \nabla \times [( \nabla  \times \mathbf{B} )  \times \mathbf{B} ]+ \eta \nabla^2 \mathbf{B}.   \label{eq:induction2}
\end{equation}
Although we do not go into details, we can simplify Equation (\ref{eq:induction2}) by properly normalizing magnetic field, time, and length (see, for example, \citealt{GaltiB03}):
\begin{equation}
    \frac{ \partial \mathbf{B} }{ \partial t} = -\nabla \times [( \nabla  \times \mathbf{B} )  \times \mathbf{B} ]+ \eta^\prime \nabla^2 \mathbf{B}.  \label{eq:eq_emhd}
\end{equation}
This is the equation for EMHD \citep{Kingsep1990review}. 
In EMHD, velocity is not an independent variable and, therefore, we do not need an equation for velocity.
We can derive velocity from magnetic field using Equation (\ref{eq:v_emhd}).

\subsection{Speed of EMHD waves}
Let us assume 
\begin{equation}
   \mathbf{B}=\mathbf{B}_0 + \mathbf{b},
\end{equation}
where we assume $\mathbf{B}_0 $ is a constant vector 
and $B_0\equiv |\mathbf{B}_0 | \gg | \mathbf{b} |$.
If we substitute this into the EMHD equation (Equation (\ref{eq:eq_emhd})), we have
\begin{equation}
    \frac{ \partial \mathbf{b} }{ \partial t} = -\nabla \times [( \nabla  \times \mathbf{b} )  \times \mathbf{B}_0 ],
    \label{eq:linear}
\end{equation}
where we ignore the second-order term and the magnetic diffusion term.
The term on the right-hand side has the form $\nabla \times ( \mathbf{j} \times \mathbf{B}_0 )$ with
$\mathbf{j} =\nabla \times \mathbf{b}$.
Since $\nabla \cdot \mathbf{j}=0$ and $\mathbf{B}_0$ is a constant vector, the vector identity
\begin{equation}
 \nabla \times ( \mathbf{j} \times \mathbf{B}_0 )
     =\mathbf{j}(\nabla \cdot \mathbf{B}_0) -\mathbf{B}_0(\nabla \cdot \mathbf{j}) 
    +(\mathbf{B}_0 \cdot \nabla) \mathbf{j} -(\mathbf{j} \cdot \nabla) \mathbf{B}_0
\end{equation}
becomes
\begin{equation}
   \nabla \times ( \mathbf{j} \times \mathbf{B}_0)
    =(\mathbf{B}_0 \cdot \nabla) \mathbf{j} = (\mathbf{B}_0 \cdot \nabla)( \nabla \times \mathbf{b})
\end{equation}
and Equation (\ref{eq:linear}) becomes
\begin{equation}
      \frac{ \partial \mathbf{b} }{ \partial t} = -(\mathbf{B}_0 \cdot \nabla) ( \nabla \times \mathbf{b}).
      \label{eq:lin_emhd}
\end{equation}

EMHD waves are actually whistler waves. A whistler is circularly polarized. In
 Fourier space, the bases that describe circular polarization are
 \begin{equation}
   \hat{\mathbf{\epsilon}}_+ \equiv ( \hat{\mathbf{s}}_1 +i\hat{\mathbf{s}}_2 )/\sqrt{2} \text{  and  }
    \hat{\mathbf{\epsilon}}_- \equiv ( \hat{\mathbf{s}}_1 -i\hat{\mathbf{s}}_2 )/\sqrt{2},
 \end{equation}
 where $\hat{\mathbf{s}}_1$ and $\hat{\mathbf{s}}_2$ are orthogonal unit vectors perpendicular
 to $\mathbf{k}$ and we assume $\hat{\mathbf{s}}_1 \times \hat{\mathbf{s}}_2 = \hat{\mathbf{k}}
 \equiv \mathbf{k}/k$.
 The $+$ waves move in the positive direction and the $-$
waves in the negative (i.e., opposite) direction with respect to $\mathbf{B}_0$.
 Let us consider a $+$ wave: 
  $\tilde{\mathbf{b}}_{\mathbf{k}} = b_0 \exp(-i \omega t) \hat{\mathbf{\epsilon}}_+$, where $b_0$ is a constant.
 In Fourier space, $\nabla \rightarrow i \mathbf{k}$ and Equation (\ref{eq:lin_emhd}) becomes
 \begin{equation}
       -i \omega \hat{\mathbf{\epsilon}}_+  
        =  -B_0 ik_{\|}  i\mathbf{k} \times  ( \hat{\mathbf{s}}_1 +i\hat{\mathbf{s}}_2 )/\sqrt{2} 
                         =- i B_0 k k_\| \hat{\mathbf{\epsilon}}_+,
 \end{equation}
 where $B_0 k_\|=\mathbf{B}_0 \cdot \mathbf{k}$ and we use 
 $\hat{\mathbf{k}} \times \hat{\mathbf{s}}_1 =\hat{\mathbf{s}}_2$ and
 $\hat{\mathbf{k}} \times \hat{\mathbf{s}}_2 =-\hat{\mathbf{s}}_1$.
 Therefore we have the dispersion relation
 \begin{equation}
    \omega = k k_\| B_0=k^2 B_0 \cos \theta,
 \end{equation}
 where $\theta$ is the angle between $\mathbf{B}_0$ and $\mathbf{k}$.
 Alfv\'en waves, which have the dispersion relation  $\omega=k_\|V_A=kV_A \cos\theta$, propagate along 
 the magnetic field at the Alfv\'en speed $V_A$ ($=B_0$ in numerical units).
 Therefore, we can conclude that a whistler wave (with a wavenumber $k$) propagates along the magnetic field at the speed of $kV_A $ ($=kB_0$ in numerical units).
 Since the propagation speed depends on the wavenumber $k$, whistler waves are dispersive.




\subsection{Critical balance in strong EMHD turbulence}
\begin{figure}[ht]
	\centering
	\includegraphics[width=0.9\textwidth]{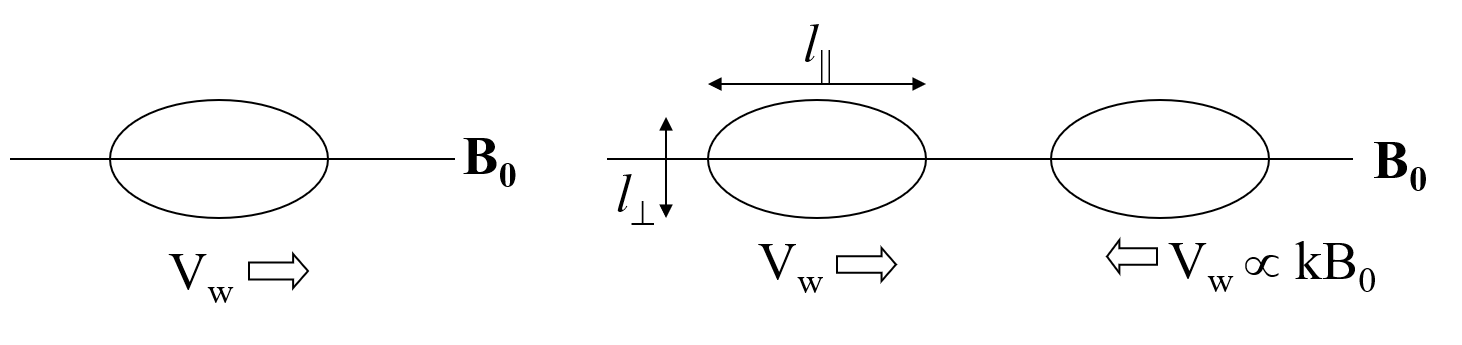}
	\caption{(Left) An EMHD wave packet (i.e., a whistler wave packet) moving in one direction. (Right) Two opposite-traveling EMHD wave packets.
      The propagation speed of an EMHD wave (i.e., whistler wave) is $kV_A$, which is equal to $kB_0$
      in the numerical units we adopt.}\label{fig:wave_emhd}
\end{figure}

As we have seen in the previous subsection, whistler waves (i.e., EMHD waves) are dispersive.
 Therefore, whistler wave packets moving in one direction can interact each other and generate turbulence.
 However, this process is relatively slow and results in inverse cascade (see \citealt{Schek09, Cho11, KimCho15})
 rather than usual forward cascade. 
 Therefore,
as in Alfv\'enic MHD turbulence, let us only consider opposite-traveling wave packets with similar sizes  (see Figure \ref{fig:wave_emhd}).
 Let us assume that their sizes in the directions parallel and perpendicular to $\mathbf{B}_0$ are $l_\|$ and $l_\bot$, respectively.
To derive scaling relations in EMHD turbulence, we can simply follow the arguments in Sections \ref{sect:cb_mhd} and \ref{sect:scale_mhd}.
We can use Figure  \ref{fig:coll} not only for MHD but also for EMHD.
Note, however, that there are two important differences between MHD and EMHD.
First, the wave propagation speed is $B_0$ in MHD and $kB_0$ in EMHD.
Second, $v_l \sim b_l$ in MHD and $v_l = \nabla \times \mathbf{b} \sim b_l/l_\bot$ in EMHD. 
 Here we assume that turbulence structures are elongated along the direction parallel to magnetic field and, hence, derivative in perpendicular direction is larger than that in parallel direction.

We assume the cross-sectional shapes of the EMHD eddies are almost circular (i.e., isotropic) before the collision (see the left panel of Figure \ref{fig:coll}).
During the collision, an EMHD eddy is distorted by shearing motion of the other EMHD eddy. 
Here, we only consider distortion in the perpendicular direction.
If velocity difference between two points, whose perpendicular separation is $l_\bot$, is $v_l$, then the distortion timescale is 
\begin{equation}
    t_{eddy} \sim l_\bot/v_l \sim l_\bot^2 / b_l.   \label{eq:eddy_emhd}
\end{equation}
On the other hand, the duration of collision is equal to the parallel size divided by the wave speed:
\begin{equation}
   t_{wave} \sim l_{\|}/(kV_A) = l_\| l_\bot / B_0,    \label{eq:wave_emhd}
\end{equation}
where we assume $k = (k_\bot^2 + k_\|^2)^{1/2} \sim k_\bot \sim 1/l_\bot$.

As in MHD, let us assume critical balance: $t_{eddy} \sim t_{wave}$.
Note that, if $t_{eddy} \sim t_{wave}$, an eddy becomes completely distorted after one collision and, therefore, one collision is enough
for energy cascade. 
From Equations (\ref{eq:eddy_emhd}) and (\ref{eq:wave_emhd}), the condition for critical balance, hence for strong EMHD turbulence, becomes
\begin{equation}
     \frac{ l_\bot }{ b_l } \approx \frac{  l_\| }{B_0},    \label{eq:critical_emhd}
\end{equation}
which is identical to the one for MHD, if the latter is expressed in terms of $b_l$ rather than $v_l$.
If turbulence driving is isotropic, the condition for critical balance at the driving scale is
\begin{equation}
    B_0  \sim b_L \sim b_{rms}.    
\end{equation}

\subsection{Scaling relations in strong EMHD turbulence}
By combining constancy of energy cascade rate and critical balance, we can derive energy spectrum and anisotropy of strong EMHD turbulence:
\begin{itemize}
     \setlength{\itemindent}{.5in}
	\item Constancy of energy cascade rate:
	 \begin{equation}
          b_l^2/t_{cas}=\text{constant},  \label{eq:cas_rate_emhd}
     \end{equation}  \\
   \item Critical balance (i.e., $t_{eddy}=t_{wave}$):
      \begin{equation}
             \frac{ l_\bot^2 }{ b_l } \sim \frac{  l_\| l_\bot }{B_0}.  \label{eq:cbal_emhd}
      \end{equation}
\end{itemize}
Since  two physically meaningful timescales, i.e., $t_{eddy}$ and $t_{wave}$, are similar (see Equation (\ref{eq:cbal_emhd})),
 we can use either of them for $t_{cas}$.
If we use $ l_\bot^2/b_l$ for $t_{cas}$, Equation (\ref{eq:cas_rate_emhd}) becomes
\begin{equation}
  b_l^3/l_\bot^2 =\text{constant,} \text{   or   }  b_l^2 \propto l_\bot^{4/3}.  \label{eq:b_l_emhd}
\end{equation}
The magnetic energy spectrum corresponding to Equation (\ref{eq:b_l_emhd}) is
\begin{equation}
  E_b(k) \propto k^{-7/3}.
\end{equation}

If we substitute Equation (\ref{eq:b_l_emhd}) into Equation (\ref{eq:cbal_emhd}), we have
\begin{equation}
   l_\| \propto l_\bot^{1/3}.   \label{eq:ani_emhd}
\end{equation}
Note that anisotropy in strong  MHD turbulence is $  l_\| \propto l_\bot^{2/3}$.
Therefore, Equation (\ref{eq:ani_emhd}) states that anisotropy in EMHD turbulence is stronger than that in MHD one.

In summary, we use constancy of energy cascade rate and critical balance to derive scaling relations in strong EMHD turbulence. 
The results shows that strong EMHD turbulence has a spectrum steeper than the Kolmogorov spectrum and anisotropy stronger than its MHD counterpart.

\subsection{Numerical tests of strong EMHD turbulence theory}

\cite{ChoL04emhd} first developed and numerically tested the theory of strong MHD turbulence.
They performed a numerical simulation of decaying EMHD turbulence.
At $t=0$, the random magnetic field is restricted to the range $ 2 \leq k \leq 4$
 in wavevector ($\mathbf{k}$) space and the amplitude of the random magnetic field is
 very close to that of the mean magnetic field (i.e., $b\sim B_0$).
 Therefore, the condition for critical balance is satisfied at $t=0$.
 
\begin{figure}[ht]
  \centering
\begin{minipage}{.5\textwidth}
	\centering
	\includegraphics[width=.94\linewidth]{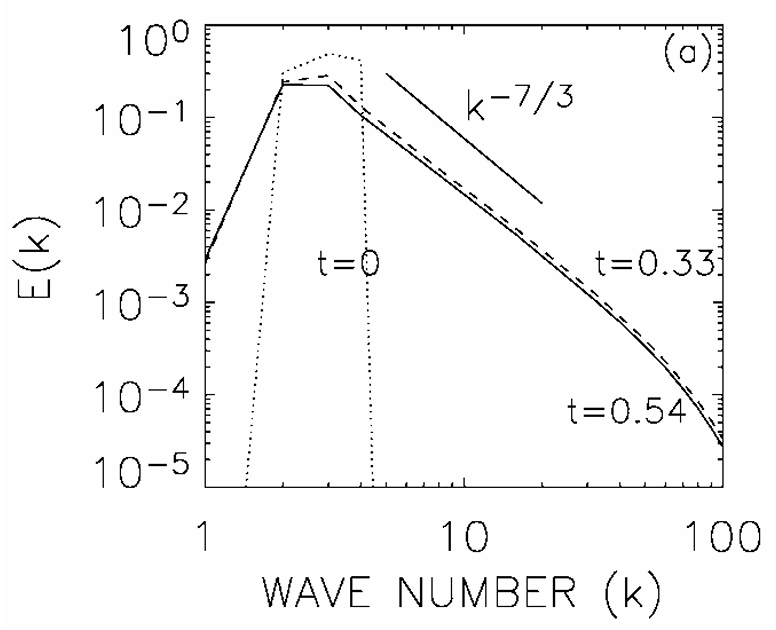}
\end{minipage}%
\begin{minipage}{.5\textwidth}
	\centering
	\includegraphics[width=.91\linewidth]{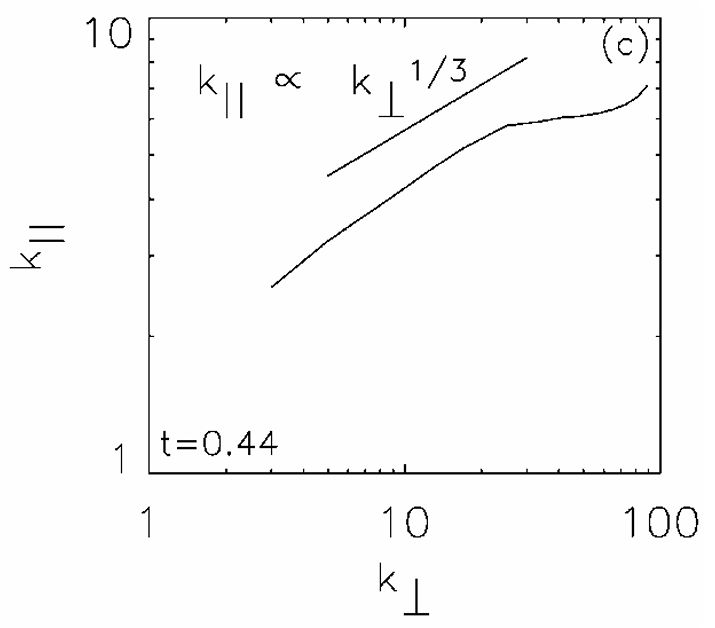}
\end{minipage}
\caption{(Left) Spectra of decaying EMHD turbulence.
      (Right) Anisotropy. From \cite{ChoL04emhd}.
}%
\label{fig:emhd}%
\end{figure}

 The left panel of Figure \ref{fig:emhd} shows that 
 the initial energy (see the dotted line) cascades down to small scales as time goes on. 
 At $t=0.33$, energy has already reached the dissipation scale (see the dashed line). 
 After energy has reached the dissipation scale, the energy spectrum decreases without changing the spectral slope
 (compare the dashed and the solid lines). 
Condition for strong turbulence is maintained for $t$'s shown in the left panel.
 The power index of energy spectrum is consistent with $-7/3$.
The right panel of Figure \ref{fig:emhd} displays anisotropy, which also confirms the theoretical prediction
$l_\| \propto l_\bot^{1/3}$.
Higher resolutions runs in \cite{ChoL09} also confirm these findings.

\section{Relativistic Alfv\'enic MHD Turbulence}  \label{sect:ffde}
 
 If magnetic field is sufficiently strong, 
  it is possible that magnetic energy density  is much larger than that of matter:
$B^2/8 /pi \gg \rho c^2$.
This may be the case in magnetosphere of a pulsar or a black hole
 (see, for example, \citealt{Goldr69, Bland77, Dunca92}) 
 or in gamma-ray bursts (see, for example, \citealt{Thomp94, Lyuti03}).
If this is the case, the Alfv\'en speed approaches the speed
 of light, and we need relativity to describe physics of
 the system. 
  If we completely ignore the energy density of matter,
 the Lorentz force vanishes and hence we can use the relativistic force-free approximation.
 As in the non-relativistic case, Alfv\'en waves
 in this extreme relativistic limit propagate at the same speed (in fact, at the speed of light) along the magnetic
 field line and hence those moving in the same direction do not
 interact.
 \cite{Thomp98} proposed a theory on strong MHD turbulence in this regime.
 According to the theory, spectrum and anisotropy of the relativistic MHD turbulence is virtually identical to
 those of its non-relativistic counterpart: $E_b \propto k^{-5/3}$ and $k_\| \propto l_\bot^{2/3}$.
\cite{Cho05} numerically confirmed the scaling relations.
 In this section, we briefly introduce relativistic force-free MHD turbulence.

 \subsection{Critical balance in relativistic  Alfv\'enic turbulence}
\begin{figure}[ht]
	\centering
	\includegraphics[width=0.8\textwidth]{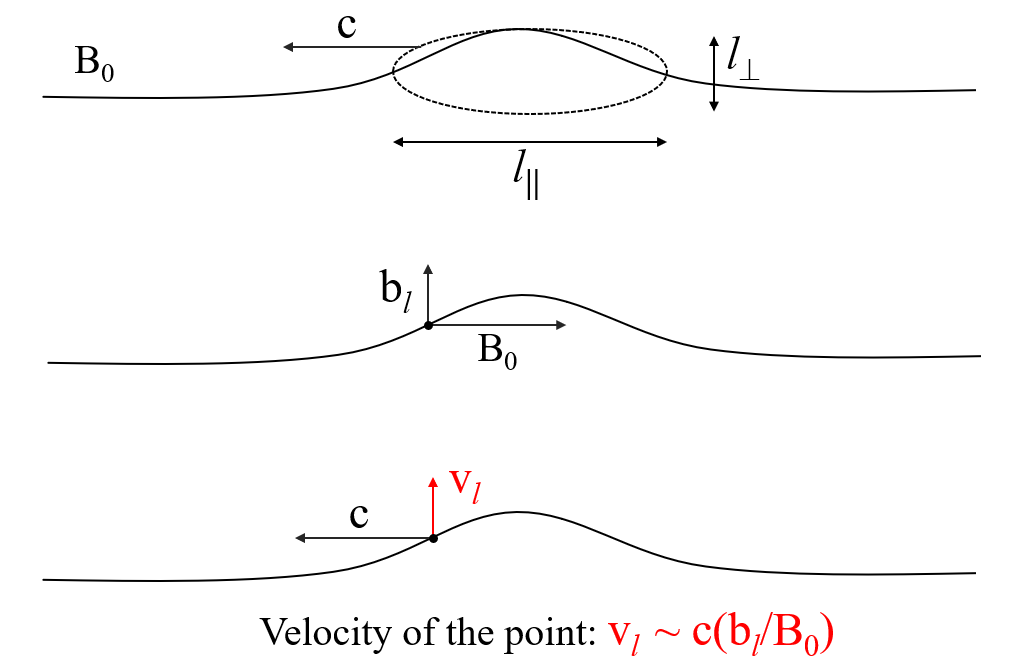}
	\caption{Propagation of an Alfv\'enic wave packet.
     (Upper) An Alfv\'enic wave packet on scale $l$ moves to the left at the speed of light.
	 (Middle) Typical `inclination' of a magnetic field line on scale $l$ is of order $\sim b_l/B_0$.
	 (Lower) Typical (perpendicular) speed of a point on the magnetic field line 
     due to propagation of the wave
    packet, is $\sim c b_l/B_0$.}\label{fig:wave_ffde}
\end{figure}

 An Alfv\'en wave propagates at the speed of light (i.e., $V_A=c$) along the magnetic field  in the relativistic force-free regime.
 Consider an Alfv\'en wave packet moving to the left (see Figure \ref{fig:wave_ffde}).
 Let  its parallel and perpendicular sizes be $l_\|$ and $l_\bot$, respectively.
 Since the strength of fluctuating magnetic field on scale the scale is $\sim b_l$, typical inclination of a magnetic field line on the scale
with respective to the mean field is $\sim b_l/B_0$ (see the middle panel of Figure \ref{fig:wave_ffde}).
When the wave packet moves to the left, a point on the magnetic field line will move up and down at the speed of
\begin{equation}
     v_l\sim c(b_l/B_0).
\end{equation}

Using the facts that $V_A =c $ and that $v_l\sim c(b_l/B_0)$, we can derive scaling relations for strong Alfv\'enic turbulence in the regime.
As in the non-relativistic MHD, let us consider
 two opposite-traveling wave packets (see Figure \ref{fig:wave}).
When they collide, they interact and generate a  turbulence cascade.
We assume the cross-sectional shapes of the relativistic Alfv\'en eddies are almost circular (i.e., isotropic) before the collision (see the left panel of Figure \ref{fig:coll}).
During the collision, the eddy is distorted by shearing motion of the other eddy. 
Here, we only consider distortion in the perpendicular direction.
Then, the distortion timescale is
\begin{equation}
    t_{eddy} \sim l_\bot/v_l \sim l_\bot / (cb_l/B_0) = l_\bot B_0/c b_l.   \label{eq:eddy_ffde}
\end{equation}
On the other hand, the duration of collision is equal to the parallel size divided by the wave speed:
\begin{equation}
   t_{wave} \sim l_{\|}/V_A = l_\| / c.    \label{eq:wave_ffde}
\end{equation}

As in non-relativistic MHD, let us assume critical balance: $t_{eddy} \sim t_{wave}$.
Note that, if $t_{eddy} \sim t_{wave}$, an eddy becomes completely distorted after one collision and, therefore, one collision is enough
for energy cascade. 
From Equations (\ref{eq:eddy_ffde}) and (\ref{eq:wave_ffde}), the condition for critical balance, hence for strong relativistic force-free MHD turbulence, becomes
\begin{equation}
     \frac{ l_\bot }{ b_l } \sim \frac{  l_\| }{B_0},    \label{eq:critical_emhd2}
\end{equation}
which is identical to the one for non-relativistic MHD, if the latter is expressed in terms of $b_l$ rather than $v_l$.
If turbulence driving is isotropic, the condition for critical balance at the driving scale is
\begin{equation}
    B_0  \approx b_L \sim b_{rms},
\end{equation}
which is also the same as its non-relativistic counterpart.

\subsection{Scaling relations in strong relativistic Alfv\'enic turbulence}
By combining constancy of energy cascade rate and critical balance, we can derive energy spectrum and anisotropy for strong relativistic force-free Alfv\'enic turbulence:
\begin{itemize}
     \setlength{\itemindent}{.5in}
	\item Constancy of energy cascade rate:
	 \begin{equation}
          b_l^2/t_{cas}=\text{constant},  \label{eq:cas_rate_ffde}
     \end{equation}  \\
   \item Critical balance (i.e., $t_{eddy}=t_{wave}$):
      \begin{equation}
             \frac{ l_\bot B_0 }{ cb_l } \approx \frac{  l_\| }{c}.  \label{eq:cbal_ffde}
      \end{equation}
\end{itemize}
We can see that these two equations are identical to those for non-relativistic MHD.
Therefore, we can conclude that strong relativistic force-free MHD turbulence has Kolmogorov magnetic spectrum
\begin{equation}
  E_b(k) \propto k^{-5/3}.
\end{equation}
and the GS95 anisotropy
\begin{equation}
   l_\| \propto l_\bot^{2/3}. 
\end{equation}

In summary, we use constancy of energy cascade rate and critical balance to derive scaling relations in strong relativistic force-free MHD turbulence. 
The results show that magnetic spectrum and anisotropy are virtually identical to those of strong non-relativistic Alfv\'enic turbulence.

\subsection{Numerical tests}
Using a numerical simulation, \cite{Cho05} confirmed that the scaling relations in the previous subsection,  which were first proposed by
\cite{Thomp98}, hold true.
We solved the following system of equations to simulate decaying relativistic force-free Alfv\'enic turbulence:
\begin{equation}
   \frac{ \partial {\bf Q} }{ \partial t }
 + \frac{ \partial {\bf F} }{ \partial x^1 }
 =0,
\end{equation}
where
\begin{eqnarray}
{\bf Q}=(S_1,S_2,S_3,B_2,B_3), \\
 {\bf F}=(T_{11},T_{12},T_{13},-E_3,E_2), \\
 T_{ij}=-(E_iE_j + B_iB_j)+\frac{ \delta_{ij} }{2} (E^2+B^2), \\
 {\bf S}={\bf E}\times {\bf B}, \\
 {\bf E}=-\frac{1}{B^2} {\bf S}\times {\bf B}.
\end{eqnarray}
Here, ${\bf E}$ is the electric field,
${\bf S}$ the Poynting flux vector and $c=1$.
(see \citealt{Komis02}).
Greek indices run from 1 to 4.
 At $t=0$ and also at later  times for which we calculated energy spectra and anisotropy, 
 the condition for critical balance is satisfied.

\begin{figure}[ht]
  \centering
\begin{minipage}{.5\textwidth}
	\centering
	\includegraphics[width=.94\linewidth]{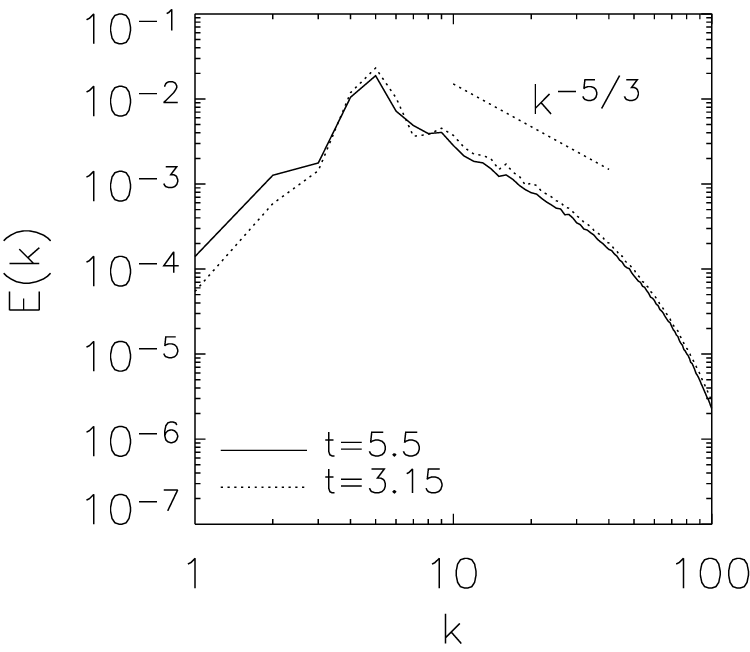}
\end{minipage}%
\begin{minipage}{.5\textwidth}
	\centering
	\includegraphics[width=.94\linewidth]{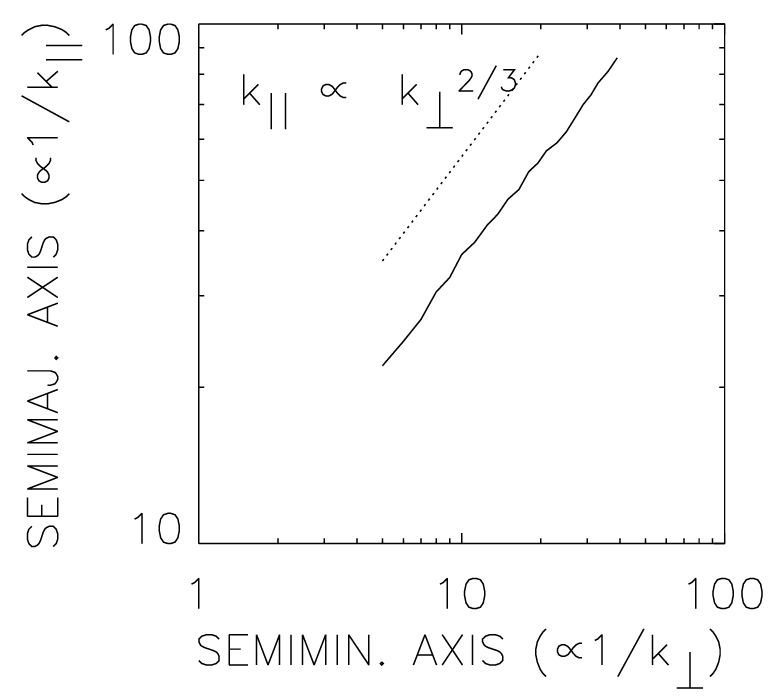}
\end{minipage}
\caption{Scaling relations in strong relativistic force-free MHD turbulence.
    (Left) Spectrum.
      (Right) Anisotropy. These scaling relations are virtually identical to those of non-relativistic strong MHD turbulence (i.e., the GS95 scaling relations).  From Cho (2005).
}%
\label{fig:ffde}%
\end{figure}

The left panel of  Figure \ref{fig:ffde} shows energy spectra of the magnetic field.
At $t=0$ (not shown in the figure)
 only large-scale (i.e., small k) Fourier modes are
 excited. At later times, turbulence develops as energy cascades down to small-scale
 (i.e., large k) modes.
 After $t \sim 3$, the energy spectrum decreases
 without changing its slope.
The spectrum at this stage is very
 close to a Kolmogorov spectrum:
 \begin{equation}
    E_b(k) \propto k^{-5/3} .
 \end{equation}
 The right panel of Figure \ref{fig:ffde} displays that anisotropy of eddy shapes follows the relation
 \begin{equation}
   l_\| \propto l_\bot^{2/3}.
 \end{equation}
These results are consistent with the theoretical prediction by \cite{Thomp98}.
\cite{Cho14ffde} showed that driven strong relativistic force-free MHD turbulence also follows the same scaling relations.

\section{MHD Turbulence in compressible medium} \label{sect:comp}
Compressible turbulence is a long-standing problem.
Since it is very difficult to analytically study scaling relations of compressible turbulence, many researchers 
rely on numerical simulations.
Nevertheless, there have been quests for scaling relations.
Some studies considered weak compressibility.
For example, \cite{ZankM93} studied weakly compressible MHD turbulence and \cite{Lithw01} proposed
theoretical predictions for compressible turbulence in strongly-magnetized high-$\beta$ plasmas, in which gas pressure is larger than magnetic pressure and 
effect of compressibility is marginal. The latter authors claimed that
Alfv\'en modes follow the GS95 scaling relations and cascade of Alfv\'en modes makes slow modes passively  follow the same scaling.
There are also studies that considered high compressibility.
For example, \cite{ChoL02prl,ChoL03mnras} studied scaling relations in supersonic MHD turbulence.
On the other hand, \cite{Krits07} studied energy cascade in supersonic hydrodynamic turbulence.
In this section, we briefly introduce the work by \cite{ChoL02prl,ChoL03mnras}, 
in which we studied `strong' MHD turbulence in highly compressible medium.
Some studies on weak MHD turbulence can be found in \cite{Chand08prl, Beatt20, Laz25}.

\subsection{Mode separation method}
\begin{figure}[ht]
	\centering
	\includegraphics[width=0.9\textwidth]{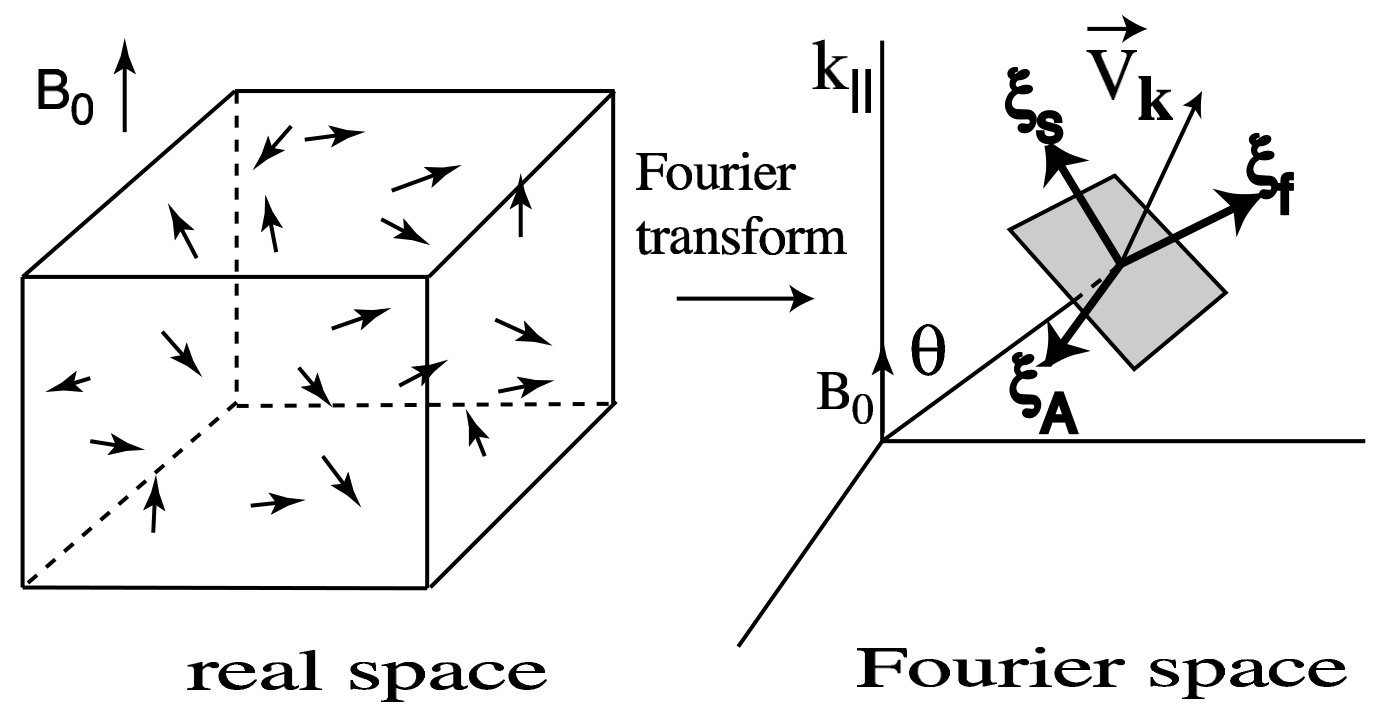}
	\caption{Separation method. We separate Alfv\'en, slow, and fast modes in Fourier
      space by projecting the velocity Fourier component ${\bf v_k}$ onto
      bases ${\bf \xi}_A$, ${\bf \xi}_s$, and ${\bf \xi}_f$, respectively. From \cite{ChoL03mnras}.}\label{fig:mode_sepa}
\end{figure}

 Three types of waves exist (Alfv\'en, slow and fast) in a compressible
 magnetized plasma.
\cite{ChoL02prl,ChoL03mnras} developed a method to separate different MHD modes and studied their scaling relations.
Figure \ref{fig:mode_sepa} explains the separation method.
Here we focus on only velocity separation.
The slow, fast, and Alfv\'en bases that denote the direction of displacement
vectors for each mode are given by
\begin{eqnarray}
   \hat{\mathbf \xi}_s \propto 
     ( -1 + \alpha - \sqrt{D} )
            k_{\|} \hat{\bf k}_{\|} 
     + 
     ( 1+\alpha - \sqrt{D} ) k_{\perp} \hat{\mathbf k}_{\perp},
  \label{eq_xis_new}
\\
   \hat{\mathbf \xi}_f \propto 
     ( -1 + \alpha + \sqrt{D} )
           k_{\|}  \hat{\mathbf k}_{\|} 
     + 
     ( 1+\alpha + \sqrt{D} ) k_{\perp} \hat{\mathbf k}_{\perp},  
   \label{eq_xif_new}
\\
 \hat{\mathbf \xi}_A = -\hat{\mathbf \varphi} 
         = \hat{\mathbf k}_{\perp} \times \hat{\mathbf k}_{\|},
\end{eqnarray}
where $D=(1+\alpha)^2-4\alpha \cos^2\theta$, $\alpha=a^2/V_A^2=\beta(\gamma/2)$,
$\theta$ is the angle between ${\mathbf k}$ and ${\mathbf B}_0$, and
$\hat{\mathbf \varphi}$ is the azimuthal basis in the spherical polar coordinate
system.
(Note that $\gamma=1$ for isothermal case.)
These bases are mutually orthogonal.

In order to separate modes,
we perform a simulation of MHD turbulence and obtain a 3-dimensional velocity data cube.
Then we compute velocity Fourier components via the Fourier transform.
We obtain Alfevn, slow, and fast Fourier velocity components by projecting the velocity Fourier component $\mathbf{v}_\mathbf{k}$ on to
$\hat{\mathbf \xi}_A$, $\hat{\mathbf \xi}_s$, and $\hat{\mathbf \xi}_f$, respectively.

\subsection{Scaling relations}

\begin{figure}[ht]
	\centering
	\includegraphics[width=0.9\textwidth]{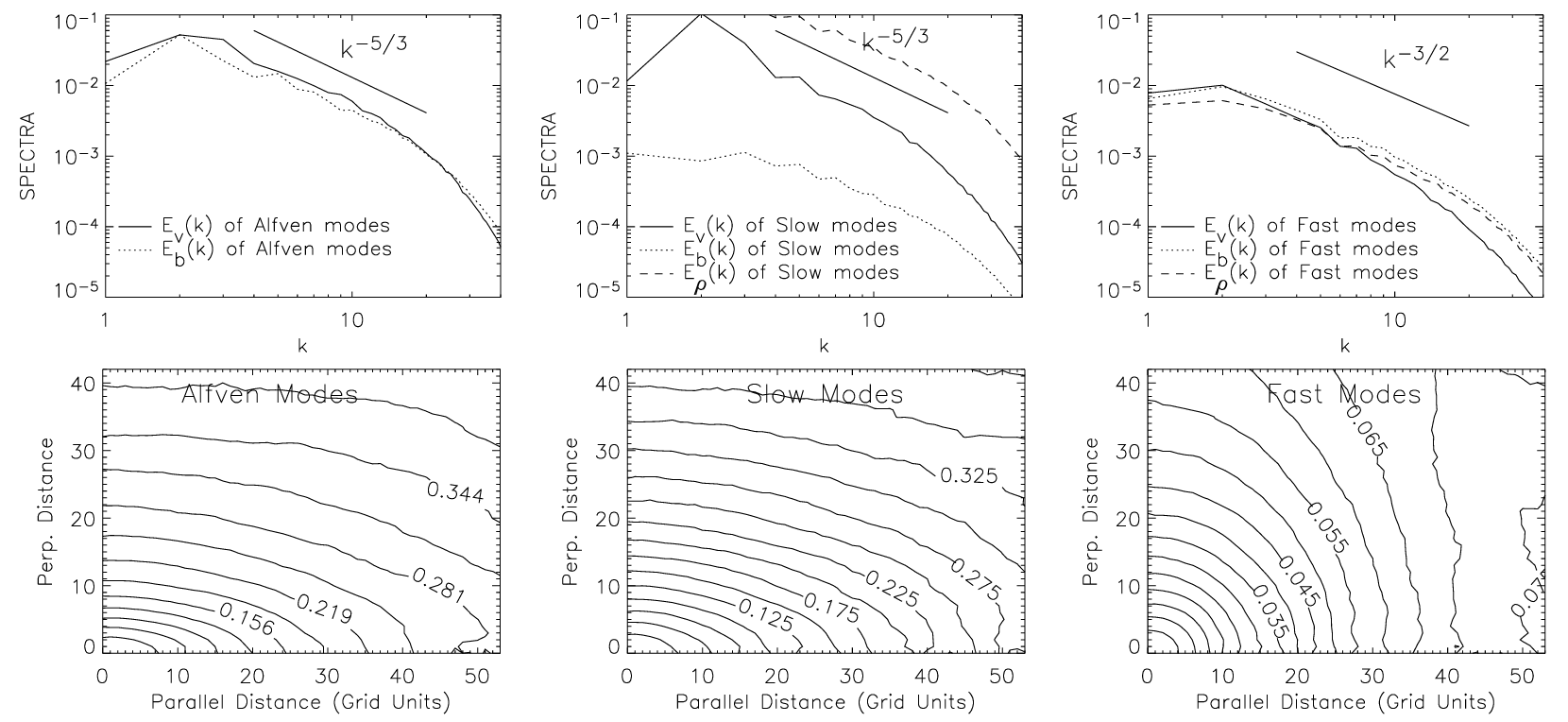}
	\caption{Scaling relations of Alfven (left panels), slow (middle panels), and fast (right panels) modes in a low $\beta$ plasma ($\beta\sim 0.2$, $M_A\sim 0.7$, and $M_s\sim 2.3$). 
         (Upper panels) From left to right, spectra for Alfv\'en, slow, and fast modes.
             Spectra of Alfv\'en and slow modes follow a Kolmogorov-like power law.
             Spectra for fast modes are shallower.
         (Lower panels) From left to right, anisotropy for Alfv\'en, slow, and fast modes.
        Contours show the second-order structure function of velocity and represent eddy shapes.
        Alfv\'en and slow modes
              follow anisotropy similar to that of GS95
             ($l_{\|}\propto l_{\perp}^{2/3}$ or $k_{\|}\propto k_{\perp}^{2/3}$).
          Fast modes does not show anisotropy. From \cite{ChoL03mnras}.
          }\label{fig:mode_spani}
\end{figure}

Figure \ref{fig:mode_spani} show the results for a supersonic low-$\beta$ plasma, where magnetic pressure is larger than gas pressure.
The average sonic Mach number is $\sim 2.3$ and critical balance is satisfied.
Upper panels display energy spectra of Alfv\'en, slow, and fast modes.
As we can see, spectra of Alfv\'en and slow modes follow a Kolmogorov spectrum, while that of fast modes looks shallower than a Kolmogorov spectrum.
Lower panels illustrate eddy shapes.
Shapes of Alfv\'en- and slow-mode eddies are scale-dependent: smaller eddies are more elongated.
Detailed calculations confirm that they follow the GS95 anisotropy.
On the other hand, fast-mode eddies do not show scale-dependent anisotropy.
Turbulence in high-$\beta$ plasmas also exhibit similar scaling relations.
Sonic Mach number also does not affect the scaling relations much.
In summary, \cite{ChoL02prl,ChoL03mnras} found that Alfv\'en and slow modes follow the GS95 scaling relations.

\section{Conclusion}\label{sec13}
In this paper, we have reviewed the scaling relations of MHD turbulence threaded by a strong mean magnetic field.
We have shown that collision of opposite-traveling wave packets is essential for generation of turbulence in various environments.
Based on critical balance, which states that one collision is enough to complete energy cascade, we have discussed following scaling relations:
\begin{itemize}
  \item strong incompressible Alfv\'enic MHD turbulence (Section \ref{sect:alf}):
     \begin{equation}
         E(k) \propto k^{-5/3} \text{  and  } l_\| \propto k_\bot^{2/3},
     \end{equation}  \\
  \item strong EMHD turbulence below the proton gyro-scale (Section \ref{sect:small}):
       \begin{equation}
    	E(k) \propto k^{-7/3} \text{  and  } l_\| \propto k_\bot^{1/3},
       \end{equation}  \\
  \item strong relativistic  Alfv\'enic MHD turbulence (Section \ref{sect:ffde}):
          \begin{equation}
         	E(k) \propto k^{-5/3} \text{  and  } l_\| \propto k_\bot^{2/3}.
         \end{equation}  \\
\end{itemize}
Alfv\'en waves or variants of Alfv\'en waves play key roles in these types of turbulence.
Note that whistler waves, which are important for EMHD turbulence, can be regarded as small-scale version of Alfv\'en waves, although the dispersion relation for the former is very different from that for the latter.

Turbulence that satisfies critical balance is called strong turbulence.
We have shown that Alfv\'en modes in strong compressible MHD turbulence exhibit the following scaling relations (Section \ref{sect:comp}):
  \begin{equation}
	E(k) \propto k^{-5/3} \text{  and  } l_\| \propto k_\bot^{2/3}.
\end{equation}  
We have also discussed scaling relations in weak incompressible MHD turbulence (Section \ref{sect:weak}):
  \begin{equation}
	E(k) \propto k^{-2} \text{  and  } l_\| =\text{constant}.
\end{equation}  
Note that weak turbulence becomes strong turbulence on scales below $\sim M_A^2 L$, where $L$ is the energy injection scale and $M_A$ is the Alfv\'en Mach number.









\section*{Conflict of interest}
The author declares no conflict of interest.

%







\bibliography{sn-bibliography}

\end{document}